\documentclass{aa}  
\usepackage[compatibility=false]{caption}
\usepackage{natbib}
\usepackage{xcolor}
\usepackage{soul}
\usepackage{graphicx}
\usepackage{txfonts}
\usepackage{amssymb}
\usepackage{amsmath}
\usepackage{mathrsfs}
\usepackage{afterpage}
\usepackage{graphicx}
\usepackage[version=4]{mhchem}
\usepackage{caption} 
\usepackage{lipsum}  
\usepackage[normalem]{ulem}
\usepackage[colorlinks=true,
    linkcolor=blue,
    citecolor=blue,
    filecolor=magenta,      
    urlcolor=cyan,
    breaklinks=true,]{hyperref}
\begin{document} 
   \title{Impact of stellar spots on the high-resolution transmission spectra of a giant planet around a Sun-like star}
   
   \author{Jennifer P. Lucero
          \inst{1,2}\fnmsep
          \and
          O. D. S. Demangeon\inst{1,2}
          \and
          E. Cristo \inst{1}
          \and
          W. Dethier \inst{1}
          \and
          N. C. Santos \inst{1,2}}

   \institute{Instituto de Astrofísica e Ciências do Espaço, Universidade do Porto, CAUP, Rua das Estrelas, 4150-762 Porto, Portugal
         \and
             Departamento de Física e Astronomia, Faculdade de Ciências, Universidade do Porto, Rua do Campo Alegre, 4169-007 Porto, Portugal\\
             }

   \date{Received October 28, 2025; accepted March 13, 2026}

  \abstract
   {Transmission spectroscopy has enabled the analysis of exoplanet atmospheres. However, a major challenge in this field is the ”noise” from
host stars, caused by stellar activity such as dark spots and bright plages. This stellar noise
can mimic or obscure signals in transmission spectra, complicating the detection and study
of exoplanetary atmospheres.}
   {We aim to characterize how unocculted stellar spots impact planetary absorption line profiles during transit by analyzing planet-occulted line distortions (POLDs), i.e. distortions originating from the non-similarity between the locally occulted stellar line and the disk-integrated stellar lines.}
   {We used the {\tt SOAPv4} tool to simulate transits of a hot Jupiter orbiting a Sun-like star under different spot configurations. We analyze the induced POLDs in the \ion{Ca}{II} K, the \ion{Na}{I} doublet, and H$\alpha$ lines.}
   {
{Our simulations show that POLDs vary with spot size, position, and stellar rotation. The \ion{Na}{I} and \ion{Ca}{II} K lines exhibit the strongest distortions, while H$\alpha$ is comparatively less affected. Low-latitude spots and higher values $v \sin i$ enhance both the amplitude and asymmetry of distortions, whereas high-latitude spots have a weaker impact. Larger spots generally lead to more pronounced modifications of line profiles, although their relative effect can decrease due to rotational broadening.}}
   {
{Our results show that non-occulted stellar spots imprint structured and line-dependent distortions in high-resolution transmission spectra, with amplitudes and velocity shifts shaped by the combined effects of activity level, stellar rotation, and spot geometry. The projected spot area emerges as the dominant factor controlling the strength of these signatures, while the line response varies, with \ion{Ca}{II} K being the most sensitive (amplitudes exceeding 2000 ppm at the highest rotation rates) and H$\alpha$ displaying distinctive asymmetric features. These findings demonstrate that stellar surface heterogeneities can mimic or alter planetary signals at the ppm level, highlighting the importance of detailed modeling for the reliable interpretation of upcoming exoplanet observations. Future work should further investigate the role of spot temperature, spot-crossing events, and more complex stellar configurations.}
}
   \titlerunning{Impact of stellar spots on the high-resolution transmission spectra of a giant planet around a Sun-like star}
\authorrunning{P. Lucero}
   \maketitle
%
\section{Introduction}
One of the key techniques used to characterize exoplanetary atmospheres is transmission spectroscopy: a powerful method that enables detailed probing of exoplanet atmospheres by observing changes in the stellar spectrum during a planetary transit. By analyzing the stellar flux at different wavelengths as the planet passes in front of its host star, one can derive the planet-to-star radius ratio as a function of wavelength \citep{2010eapp.book.....S}. This technique enables the detection and characterization of chemical species in exoplanet atmospheres, providing critical insights into the nature of distant worlds.

Using this technique, a wide range of species have been detected in exoplanetary atmospheres using both space- and ground-based telescopes, such as Hubble, Spitzer, and James Webb in space and VLT, Keck, and GTC on the ground. For example, \ce{TiO} has been identified in the atmosphere of the hot Jupiter WASP-19b \citep{Sedaghati_2017}, and high-resolution transmission spectroscopy has revealed \ce{CO} and \ce{CH4} in HD 209458b (\citealt{Snellen_2010}, \citealt{Giacobbe_2021}). Furthermore, \ion{Li}{I}, \ion{Na}{I} and \ion{Fe}{I} in WASP-76b \citep{tabernero, Azevedo_Silva_2022}, as well as the presence of \ion{Ni}{I}, \ion{Ti}{I} and \ion{Cr}{I} in KELT-9b \citep{hoeijmakers}, demonstrate the growing power of this technique.

Transmission spectroscopy critically depends on an accurate understanding of the wavelength-dependent brightness of the stellar disk being occulted. However, stellar surface inhomogeneities, such as spots, faculae, plages, and granulation, along with global stellar phenomena such as oscillations, rotation, and center-to-limb variations pose major challenges. These features distort or shift the stellar spectrum and, when unocculted during transit, can alter the observed spectra by mimicking broadband signals (e.g., Rayleigh scattering slopes) or modulating the amplitude of narrow-band atomic and molecular lines \citep{berdyugina,oshag, mallon, Boldt_2020}. This contamination, commonly referred to as the Transit Light Source Effect (TLSE), arises from stellar surface features during planetary transits. Although it can occur in all stars, it is most often studied in F to M-type dwarf stars, as the majority known transiting exoplanets orbit these stars \citep[e.g.,][]{Rackham_2018}. It originates from the discrepancy between the disk-integrated stellar spectrum and the spectrum of the transit chord.

A well-studied example is the K-dwarf HD 189733, which hosts the transiting hot Jupiter HD 189733b \citep{bouchy} and has become a benchmark system for transmission spectroscopy due to its large atmospheric scale height and its bright, magnetically active host star (e.g., \citealp{pont, sing, cristo}). Early low-resolution studies showed that unocculted spots and plages affect broadband transmission spectra (\citealp{pont2013}; \citealp{oshag}). However, at high-resolution, these broadband variations are removed through normalization, and the dominant source of contamination arises instead from line-profile distortions.

This distinction is illustrated by \ion{Na}{I} D observations of HD 189733b \citep{Wyttenbach, keles, mounzer} reported deep, spectrally resolved absorption in the transmission spectrum, initially analyzing the system under the assumption that stellar effects such as the Rossiter–McLaughlin effect and center-to-limb variation (CLV) would largely average out over the transit. Subsequent studies demonstrated that this assumption is not valid,  \citet{Louden_2015} and \citet{Casasayas_Barris_2017} showed that the combined impact of the RM effect and CLV variation distorts stellar line profiles during transit, biasing both the measured \ion{Na}{I} absorption (depth and shift), leading to a strong impact on the inferred wind speed.

Molecular detections at high-resolution require strict validation, as the dominant sources of spurious signals differ from those in low-resolution data. This phenomenon has notably complicated CO detection claims for systems such as HD 189733 b \citep{brogi}. For 51 Peg b, \citet{Brogi_2013} showed a significant temporal inconsistency, with signals disappearing later in the observation hours. In addition, infrared transmission studies of WASP-80 b showed how stellar molecular line distortions bias the inferred planetary signal \citep{Carleo_2022, brogi}.

High-resolution optical transmission spectra are particularly susceptible to contamination from stellar activity \citep{Cauley_2018}. In particular, stellar spots and plages can imprint spurious signals onto the spectra, especially near lines such as H$\alpha$ and \ion{Na}{I} D. Even low-contrast active regions can produce signals resembling planetary absorption, particularly in stars exhibiting significant chromospheric variability. These effects underscore the importance of accounting for stellar activity when interpreting narrowband features in high-resolution spectra, such as H$\alpha$ and \ion{Na}{I} D.

The success of upcoming missions and instruments, aimed at detecting Earth-like planets around Sun-like stars, hinges on a deep understanding of stellar noise. The present methods fail to achieve the precision required to separate stellar contamination from true planetary signatures. The search for life-sustaining planets is becoming a central focus of exoplanet research efforts \citep[e.g., ANDES project for the ELT;][]{marconi}, further highlighting the urgency of overcoming these limitations. Furthermore, no previous studies have systematically addressed this issue using high-resolution transmission spectroscopy, as provided by instruments such as ESPRESSO \citep{espresso}. This shortcoming limits our ability to unambiguously identify Earth-like exoplanets and assess their atmospheres.

In this paper, we investigate the effects of non-occulted stellar spots on high-resolution transmission spectra of exoplanets. We focus on a Sun-like star hosting a hot Jupiter and examine spot-induced signatures across key spectral regions, including \ion{Ca}{II} K (6 ~\AA), the \ion{Na}{I} D doublet (5889.95  ~\AA\ and 5895.92 ~\AA), and H$\alpha$ (6562.85~\AA). These lines have been detected several times through these specific absorption lines in different exoplanets. Moreover, as they are deep lines, they are known to have strong Planet-Occulted Line Distortions (POLDs), thus making them a good candidate for our study. The methodology of our simulations is described in Section~2, followed by the presentation of the results in Section~3. Section~4 provides a detailed discussion and interpretation of the findings, and Section~5 concludes with a summary and key findings.

\section{Description of the model} 
\label{sec2}
\begin{figure}[h]
\begin{center}
\includegraphics[scale=0.61]{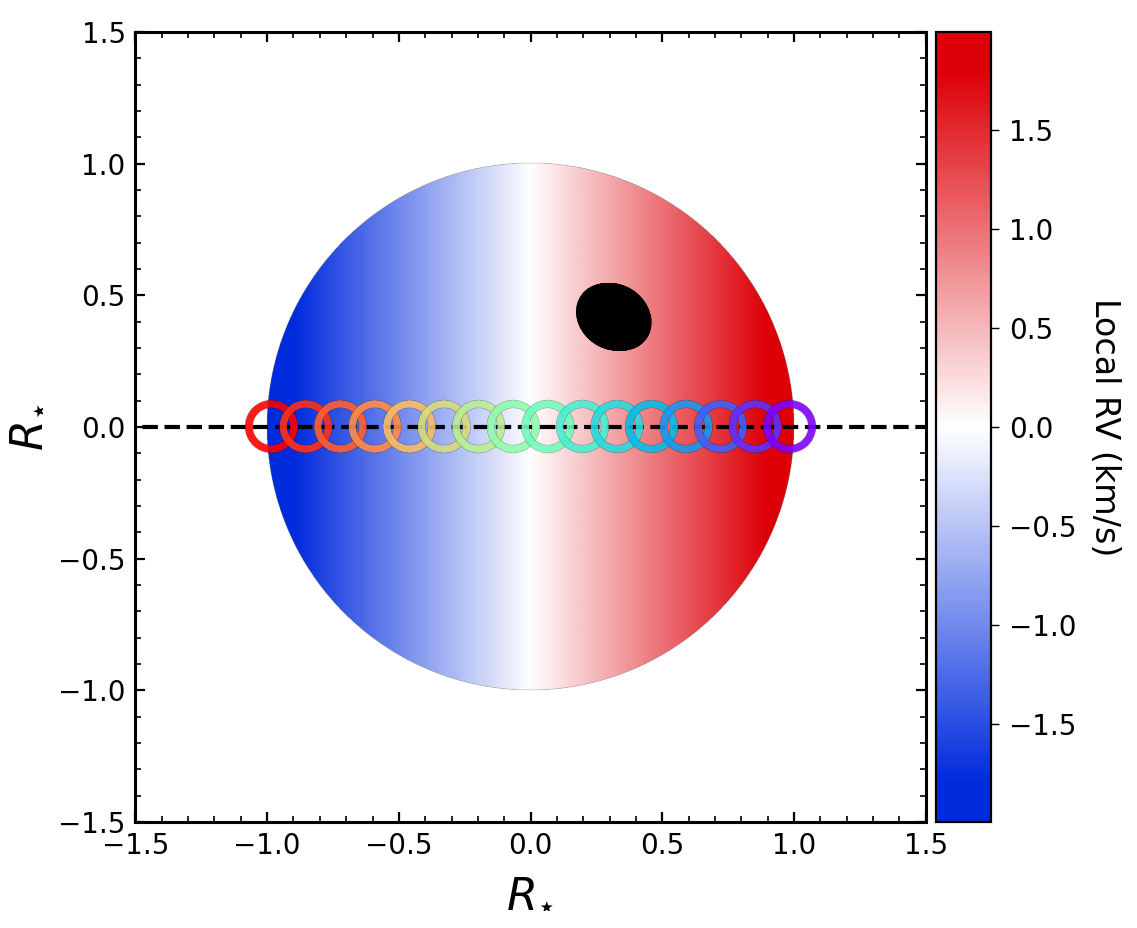}
\caption{Doppler map of the visible hemisphere of the Sun-like star. Colored open circles indicate the planet positions during transit, moving from red to purple as the transit progresses. For clarity, the data points are undersampled in the figure; the full dataset was used in the analysis. The dotted black line shows the transit path, while the black disk represents a simulated spot with a filling factor of 1\% smeared over time.}
\label{fig1}
\end{center}
\end{figure}

To evaluate how stellar spots affect high-resolution transmission spectra, we performed simulations using the {\tt SOAPv4} forward modeling code \citep{Cristo_2025}, an updated and extended version of the {\tt SOAP} tool \citep{boisse2012, soap2013, Dumusque_2014, gpu}. {\tt SOAPv4} is a tool that simulates stellar spectra during planetary transits, including the effects of surface inhomogeneities such as spots and plages. Figure~\ref{fig1} illustrates a representative case of a Jupiter-sized planet transiting a Sun-like star with a spot in the visible hemisphere. Using this framework, our objective is to isolate and characterize the spectral imprint of non-occulted stellar spots under controlled conditions.

While several tools have been developed to simulate stellar surface heterogeneities and account for the direct and indirect effects of stellar activity on transit light curves, such as variations in flux and spectral line distortions, these are often tailored to photometric corrections or broadband analysis. Examples include {\tt StarSim} \citep{Herrero2016}, {\tt PRISM} \citep{Prism}, {\tt Ksint} \citep{Montalto}, {\tt PyTranSpot} \citep{juvan}, and {\tt SAGE} \citep{Chakraborty_2024}. 
In contrast, {\tt SOAPv4} provides flexible transit modeling and supports both synthetic and observed spectra, making it particularly suitable for high-resolution transmission spectroscopy studies.
\begin{figure*}[h]
\begin{center}
\includegraphics[scale=0.52]{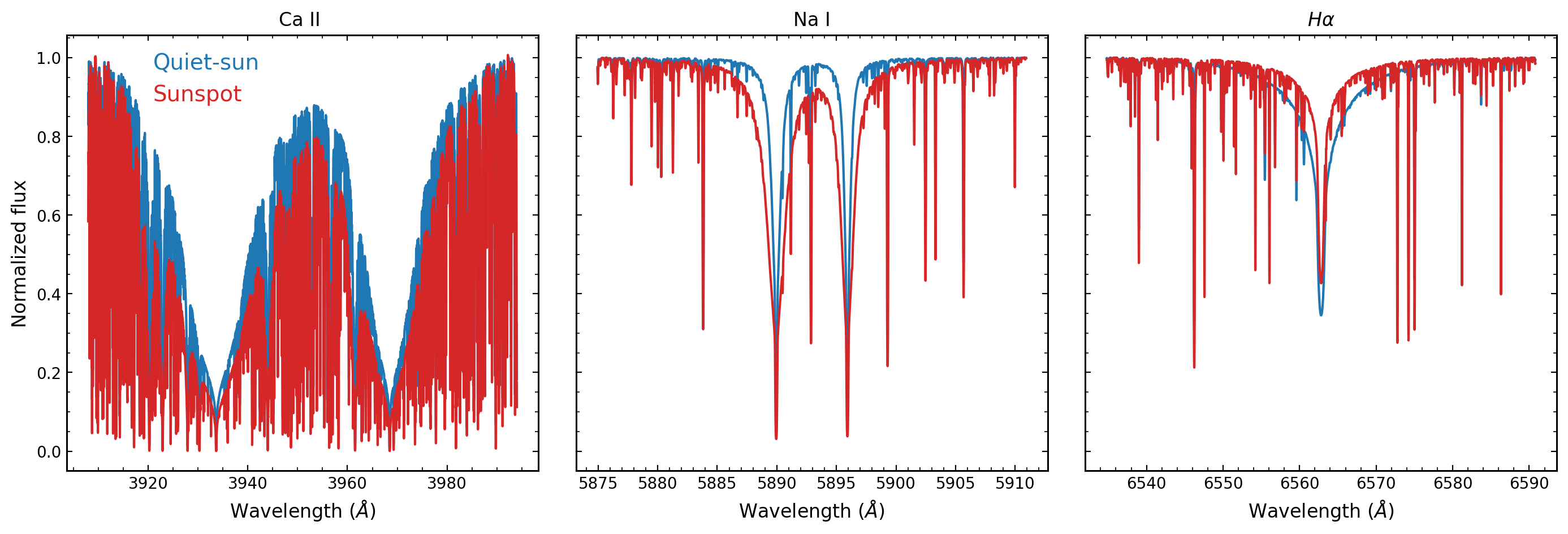}
\caption{Normalized simulated spectra of the quiet Sun (blue) and of a spot (red) for \ion{Ca}{II}, \ion{Na}{I} and H$\alpha$ spectral regions. }
\label{fig2}
\end{center}
\end{figure*}\\ \\
{\tt SOAPv4} can simulate stellar spectra using stellar atmospheric models, such as PHOENIX \citep{phoenix}, or observed spectra, such as those obtained with the Fourier Transform Spectrometer (FTS) and provided in the Kitt Peak Solar Atlas \citep{ftspec}. In this work, we restrict ourselves to the use of PHEONIX models.
PHOENIX provides high-resolution spectra across a broad wavelength range and covers a wide grid of stellar parameters, including effective temperature, surface gravity, and metallicity. These models assume LTE conditions and describe only the stellar photosphere, with no chromospheric contribution, which is a limitation of this study.
\subsection{Transit simulations}
\label{sec2.1}
To initialize a simulation in {\tt SOAPv4}, the physical parameters of both the quiet star and the active regions should be specified, including metallicity [Fe/H], surface gravity $\log g$, effective temperature $T_{\text{eff}}$, and the temperature contrast between the quiet star and the active region $\Delta T_{\text{spot}}$. We also defined the pixel scale, which sets how finely the stellar disk is discretized. In practice, the disk is projected onto a grid of $N \times N$ pixels, where each pixel represents a cell of the stellar surface with its own local spectrum. For each component (photosphere and active region), {\tt SOAPv4} automatically retrieves the synthetic spectra from the {\tt PHOENIX} precomputed grid that most closely match the provided stellar parameters, ensuring realistic local spectra across the stellar disk. The default spectral resolution in {\tt SOAPv4}, corresponds to the HARPS spectrograph (\(R \sim 115{,}000\); \citealt{2003Mayor}), and is applied through convolution with a Gaussian instrumental profile, whose FWHM corresponds to $\lambda$/\(R\), where \(R\) is the user-specified resolution \citep{boisse2012}. The wavelength grid on which the spectra are evaluated is fully user-defined and independent of \(R\).

In this work, we focus on a solar-like star. The solar atmosphere is known to be highly inhomogeneous, with most of the variability in the photosphere and chromosphere driven by magnetic activity, including spots, faculae, and plages. Although such features cannot be spatially resolved in other stars, previous studies suggest that similar magnetic phenomena are present across a wide range of stellar types \citep[e.g.,][]{berdyugina, Reiners_2012}. 

In our simulation, we adopt solar values for the stellar parameters (see Table~\ref{tab1}). The darkening of the limb is described using a quadratic law with coefficients $u_1$ and $u_2$ calculated for each spectral region shown in Figure \ref{fig2} using the darkening toolkit of the limb ({\tt LDTk}; \citealt{LDTK}).

Regarding the planetary companion, we consider the orbital period P, semi-major axis (\(a\)), orbital inclination (\(i_{p}\)), and eccentricity (\(e\)) as in Table 1. The physical properties of the planet include its mass (\(M_p\)) and radius (\(R_p\)), for the hot Jupiter-like case. The planetary radial velocity semi-amplitude (\(K_p\)) is also considered, calculated from $M_p$, $M_\star$, P, and a.
All stellar and planetary input parameters used in the simulations are listed in Table~\ref{tab1}.
\begin{table}[ht!]
      \caption[]{Input stellar and planetary parameters.}
         \label{tab1}
     $$ 
         \begin{array}{p{0.5\linewidth}l}
            \hline
            \noalign{\smallskip}
            Parameter description     &  \mathrm{Value} \\
            \noalign{\smallskip}
            \hline
            \noalign{\smallskip}
            Stellar Information & \\
            $T_{\mathrm{eff}}$  (K) & 5778\hspace{2mm}   \\
            log(\textit{g}) $\mathrm{(dex)}$ & 4.5\hspace{2mm}  \\
            $\mathrm{[Fe/H]}$  & 0             \\
            \textit{$i_{\star}$}[$^\circ$]& 90\\
            $u_1$ - \ion{Ca}{II} K & 0.8496\\
            $u_2$ - \ion{Ca}{II} K & -0.0591\\
            $u_1$ - \ion{Na}{I} D & 0.5296\\
            $u_2$ - \ion{Na}{I} D & 0.1344\\
            $u_1$ - H$\alpha$ & 0.3864\\
            $u_2$ - H$\alpha$ & 0.1854\\
           $M_\star\ (M_\odot)$ &1 \\
           $R_\star\ (R_\odot)$& 1\\

           \noalign{\smallskip}
            \hline
            \noalign{\smallskip}
            Planet Information    &  \\
            Period [days] & 3 \\
            a $(R_\odot)$ & 10.1 \\
            \textit{$i_{p}$}[$^\circ$] & 0 \\
            \textit{e} & 0 \\
            $M_p\ (M_\oplus)$ & 318\\
            $R_p/R_\odot$ (hot-Jupiter) & 0.1\\
            $K_p\ (\mathrm{km\ s^{-1}})$ & 137.44 \\
            \noalign{\smallskip}
            \hline
         \end{array}
     $$ 
   \end{table}
\begin{figure*}[ht!]
\begin{center}
\includegraphics[scale=0.45]{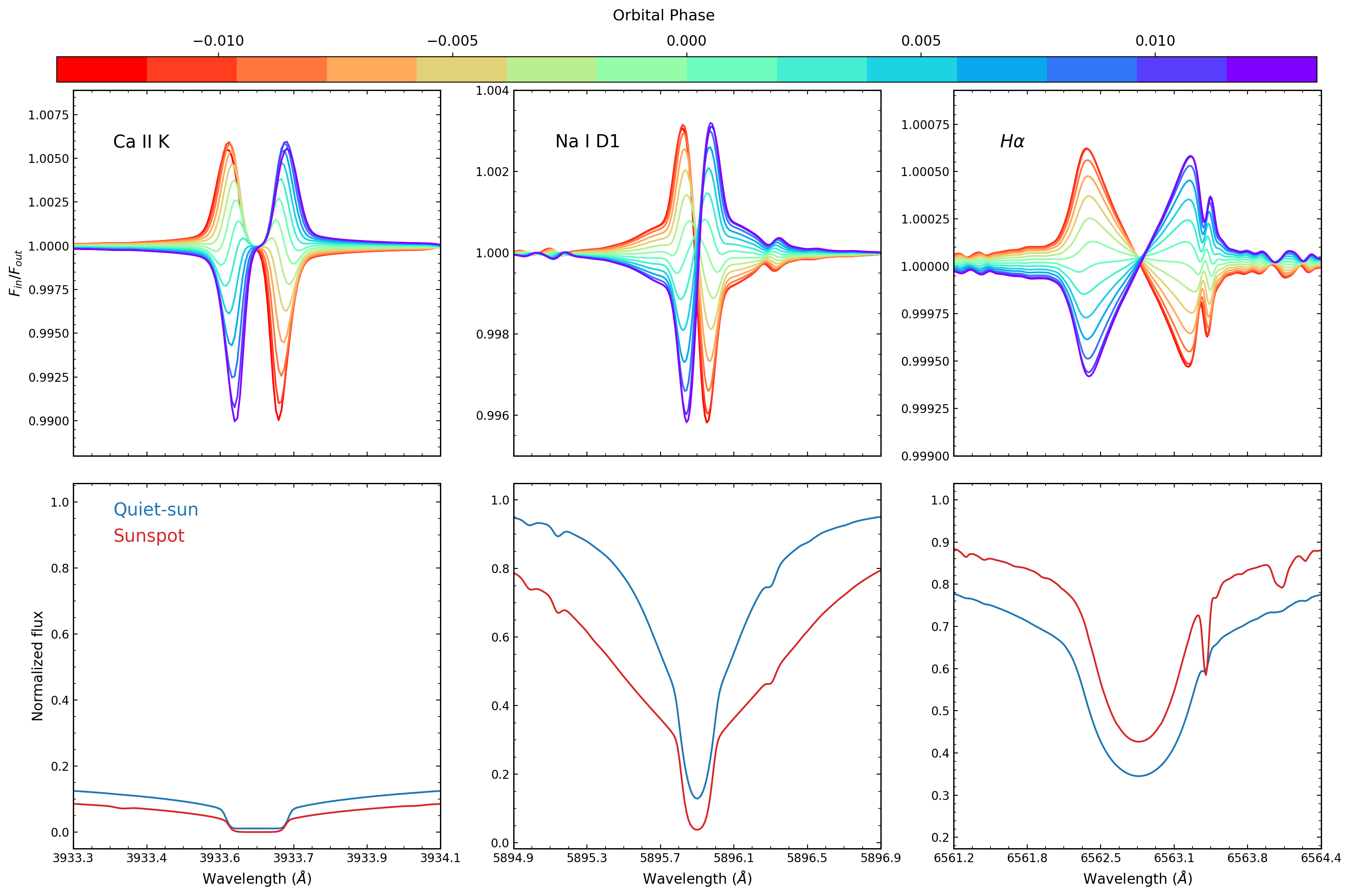}
\caption{Top Panel: Transmission spectrum in the stellar rest frame for a planet-spot configuration with planet orbital phase (position) indicated by the colored circles. The colour coding identical to the one use in Figure \ref{fig1}. Bottom Panel: Zoom of the normalized spot and quiet Suns pectra shown in Figure \ref{fig2}  for the three spectral regions studied.}
\label{fig3}
\end{center}
\end{figure*}

Stellar spots are modeled as circular features defined at the disk center. Their apparent shape is then projected following spherical coordinates from the disk center to the limb. We restrict our analysis to non-occulted spots. In order to control the temperature contrast between the spot and the quiet photosphere, {\tt SOAPv4} normalizes the PHOENIX spectra of both the spot and the quiet Sun. To do so it estimates the continuum by analyzing the first and last 10\% of the wavelength range considered. It computes the flux beyond the 99th percentile in each region and fits a first-degree polynomial across the range. The flux array is then divided by this polynomial fit. We focused on three key spectral regions: the \ion{Ca}{II} K line, covering 3907.66--3994.47~\AA, the \ion{Na}{I} D doublet, from 5874.95 to 5910.92~\AA, and H$\alpha$, spanning 6534.80--6590.80~\AA. Figure~\ref{fig2} illustrates the normalized spectra of the quiet star and a spot for these selected spectral regions.
For the temperature contrast of the spot ($\Delta T_{\text{spot}}$), we adopted a fixed value of $-663\,\mathrm{K}$, consistent with the average difference between sunspots and the quiet solar photosphere \citep{meunier}, and in agreement with sunspot spectral atlases \citep{ftspot}. While this value is smaller than the typical contrasts of $\sim1000$--$1800\,\mathrm{K}$ reported for active regions in G- and K-type stars \citep{berdyugina}, the Sun represents a well-studied case where the measured contrasts are closer to $\sim600$--$700\,\mathrm{K}$. We therefore adopted the solar value as a reference baseline, as the present work focuses primarily on the geometric configuration of stellar spots (size, position, and distribution), rather than on the dependence of the results on the temperature contrast. We do not explore variations in $\Delta T_{\text{spot}}$. However, previous work has shown that the contrast effect from spots is small for atomic lines, independent of the value of $\Delta T_{\text{spot}}$ \citep{Cauley_2018}.

For our study, we varied (1) the fractional spot coverage \(f_{\mathrm{sp}}\), (2) the spot’s latitude and longitude on the stellar disk, and (3) the stellar projected rotational velocity \(v \sin i\). The adopted parameter space is summarized in Table \ref{t2}.

For spot coverage, we modeled a single circular spot and explored values from 0.1\% to 1\%, consistent with minimum and maximum solar activity levels \citep{shapiro, apai2018understanding}. Additionally, we included a higher-activity scenario with 3\% coverage to probe more extreme configurations. We used a spatial grid of \(1300 \times 1300\) pixels for the stellar disk, which allowed small spots to be resolved across multiple cells, avoiding undersampling. 

\begin{figure*}[h]
\begin{center}   
\includegraphics[scale=0.58]{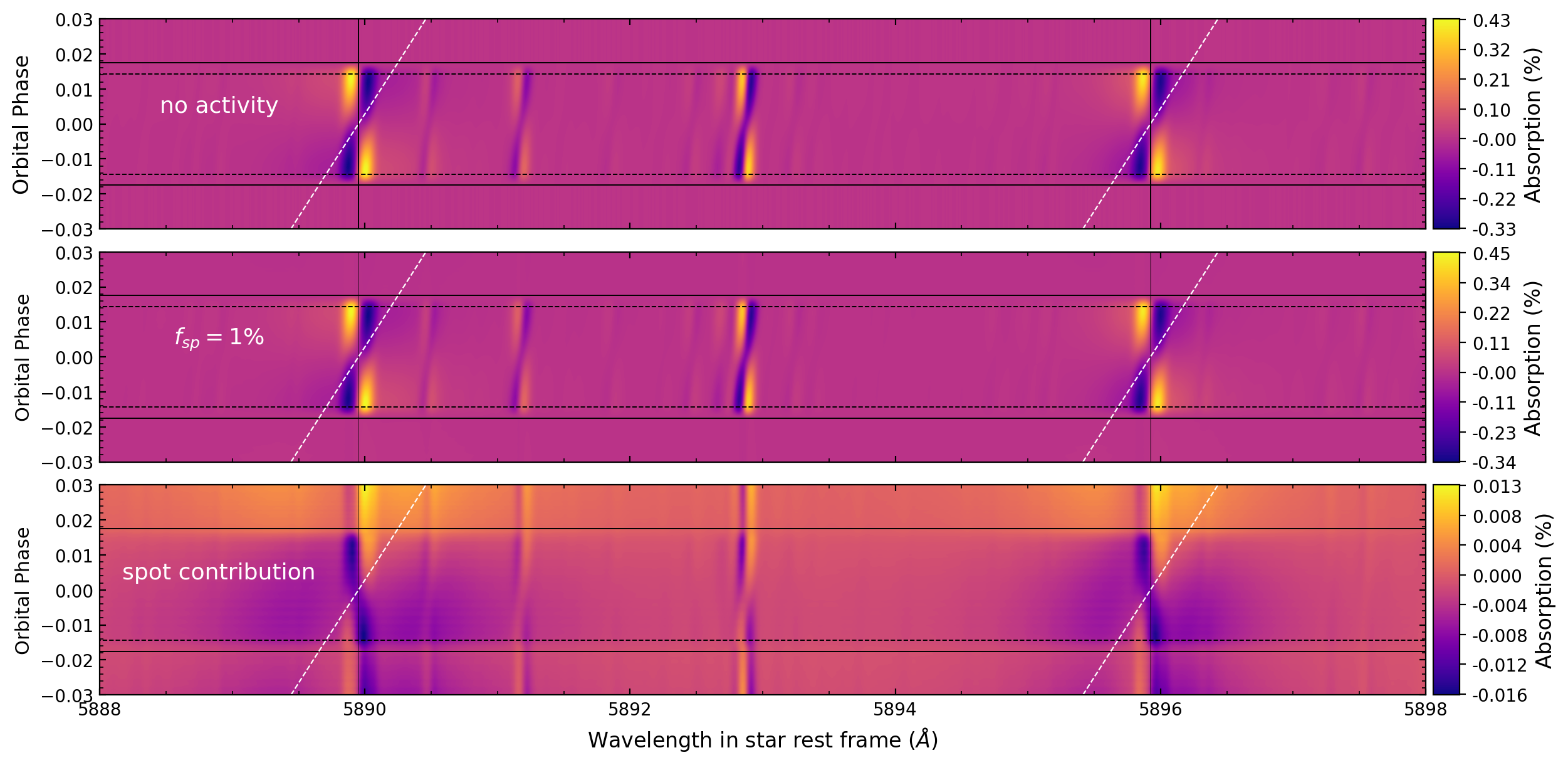}
\caption{Tomography plots of the individual absorption spectra around the Na\,\textsc{i} doublet for a simulated hot Jupiter. The top panel: shows the case without stellar activity, the middle panel includes the effect of a stellar spot with a coverage fraction of \( f_{\mathrm{sp}} = 1\% \), and the bottom panel displays the isolated spot contribution, obtained as the difference between the two upper panels. The color scale represents the absorption \((1 - F_{\mathrm{in}} / F_{\mathrm{out}})\) in percent, while the dashed white line traces the expected planetary trail. The horizontal black lines indicate the four contact points of the transit. For this case we adopt \(v \sin i = 2\,\mathrm{km\,s^{-1}}\).}
\label{fig4}
\end{center}
\end{figure*}

\begin{table}[ht!]
    \centering
    \caption{Model parameters and explored values}
    \label{t2}
    \begin{tabular}{p{0.5\linewidth} l l}
        \hline
        \noalign{\smallskip}
        Parameter description & Symbol & Values \\
        \noalign{\smallskip}
        \hline
        \noalign{\smallskip}
        Spot coverage fraction & $\textit{f}_{\mathrm{sp}}$ & 0.1, 1,   3\% \\
        Latitude & $\theta_{\mathrm{lat}}$ & 25, 50, 80$^\circ$ \\
        Longitude & $\theta_{\mathrm{lon}}$ & $\pm$  20, 50, 80$^\circ$  \\
        Sun rotational velocity & $v\sin i$ & 2, 6, 10 $\mathrm{km\ s^{-1}}$ \\
        Spot temperature difference & $\Delta T_{\text{spot}}$ & $-663$ K \\
        Sun effective temperature & $T_{\text{eff}}$ & 5778 K \\
        \noalign{\smallskip}
        \hline
    \end{tabular}
\end{table}

For the smallest spot simulated (if located in the center of the stellar disk) this configuration yields a resolution of approximately 2656 grid cells covering the spot area by using the Equation \ref{eq1} and the spatial grid previously mentioned. This number vary slightly during the transit due to the relative motion of the spot across the disk and the projection effects at different latitudes, particularly at high viewing angles. 
In {\tt SOAPv4}, spot size is defined as:
\begin{equation}
\label{eq1}
\frac{R_{\text{spot}}}{R_{\star}} = \sqrt{\frac{2A_{\text{spot}}}{A_{\star}}} = \sqrt{\frac{2f_{\text{sp}}}{100}}
\end{equation}
where \(A_{\text{spot}}\) and \(A_{\star}\) are the spot and stellar surface areas, respectively. This expression ensures that the desired fractional coverage is preserved for circular spots.

We varied spot latitudes and longitudes within the visible hemisphere, ensuring that the transit chord did not intersect the spot. For all of our simulations, the transit impact parameter was fixed at \textit{b} = 0, so that the planet always crosses the stellar equator. In practice, whether a spot is occulted or not also depends on the planetary radius: a larger planet could still cover features located at relatively low latitudes. To avoid this, we restricted the spot locations to latitudes sufficiently far from the transit chord, ensuring that the contamination we measure arises exclusively from non-occulted active regions. While this work focuses on unocculted active regions, spot-crossing events can generate localized distortions in both the amplitude and velocity structure of the transmission spectrum, depending on the position and size of the spot and the stellar rotation velocity \citep[e.g.,][]{Cristo_2025}. Although the Sun rotates at roughly 
$\sim$ $2\,\mathrm{km\,s^{-1}}$ \citep{Dumusque_2014}, we expanded the range of \(v \sin i\) values up to 10\, $\mathrm{km\,s^{-1}}$ to cover typical values found in G- and K-type planet-hosting stars \citep{Reiners_2012, Cauley_2018} and assess whether stellar rotation enhances the spectral impact of active regions. 
\subsection{Computation of the transmission spectrum}
\label{sec2.2}
In our simulations, we computed the integrated stellar spectrum at 100 time steps during the planetary transit. This sampling provides sufficient temporal resolution to capture the spectral variations across the transit while maintaining computational efficiency. We constructed the transmission spectrum ($F_{\mathrm{in}}/F_{\mathrm{out}}$) for each spectral region by dividing each in-transit spectrum ($F_{\mathrm{in}}$) by a master out-of-transit spectrum ($F_{\mathrm{out}}$) in the stellar rest frame. Here, $F_{\mathrm{in}}$ refers to the stellar spectrum obtained between the second and third contacts of the transit (T2-T3, i.e., the full transit phase), while $F_{\mathrm{out}}$ denotes the reference stellar spectrum outside of transit, constructed as the average (master) of multiple out-of-transit exposures. Figure~\ref{fig3} shows an example of one of our simulated transmission spectrum for the different spectral regions in the presence of a spot in the stellar rest frame.

We followed a similar procedure used in {\tt SOAPv4} to express the spectra in the planetary rest frame. Using the known keplerian velocity and the orbital phases of the planet, we then applied a Doppler shift to the spectra via cubic interpolation, thereby converting them into the planetary rest frame. In this case, we did not correct the spectra for the slope induced in the radial velocities by the spot, as the simulations are defined in the true stellar rest frame. In observations, such slopes are generally not identified or corrected explicitly, since the measured velocities already include the contribution from stellar activity. Moreover, Section \ref{sec4.1} discusses how the results change when this slope correction is included, allowing for a direct comparison between both approaches.
Once the spectra are in the planetary rest-frame, we computed the mean in-transit absorption spectrum for each configuration. See details in Section \ref{sec3}.
\subsection{Definition of spectral metrics}
\label{sec2.3}
In a previous study, \citealt{Chakraborty_2024} investigated the impact of unocculted active regions on the observed and true transit depths at different wavelengths, using low-resolution simulations (see Equations 3 and 6). Following their approach, we extend the analysis to the high-resolution domain, aiming to assess how stellar spots of different sizes and positions affect the shape of spectral lines. To quantify these effects, we employ two complementary metrics designed to capture line distortions.
First, we measured the amplitude of the spot contribution by calculating the difference between the maximum and minimum within 1~$\AA$ on each side of the line center, to consistently include the full extent of the spectral features.
\begin{figure*}
\begin{center}
\includegraphics[scale=0.43]{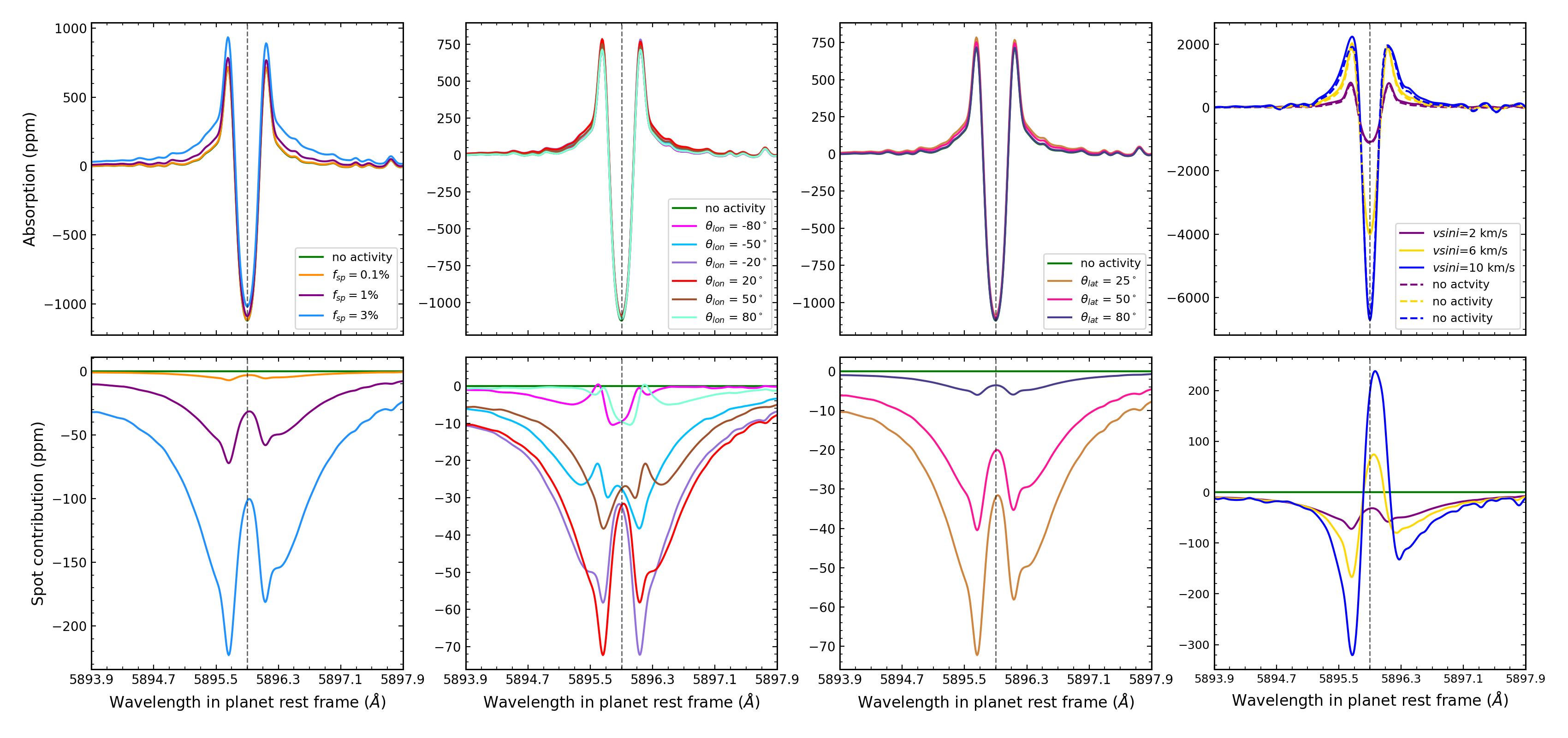}
    \caption{Top: Mean in-transit absorption spectra in the planetary rest frame around the \ion{Na}{I} D1 line. Bottom: Spot contribution as the difference obtained by subtracting the quiet-star profiles from the star with a spot profiles. Each column represents (from left to right) the impact of spot size, spot position (longitude and latitude, respectively), and stellar rotational velocity. According to each case the fixed parameters were \( f_{\text{sp}} \) = 1\%, $\theta_{lat}$ = $25^\circ$, $\theta_{lon}$ = $20^\circ$ and \(v \sin i\) = $2\,\mathrm{km\,s^{-1}}$.} 
    \label{fig5}
\end{center}
\end{figure*}
\begin{figure*}[ht!]
\begin{center}
\includegraphics[scale=0.43]{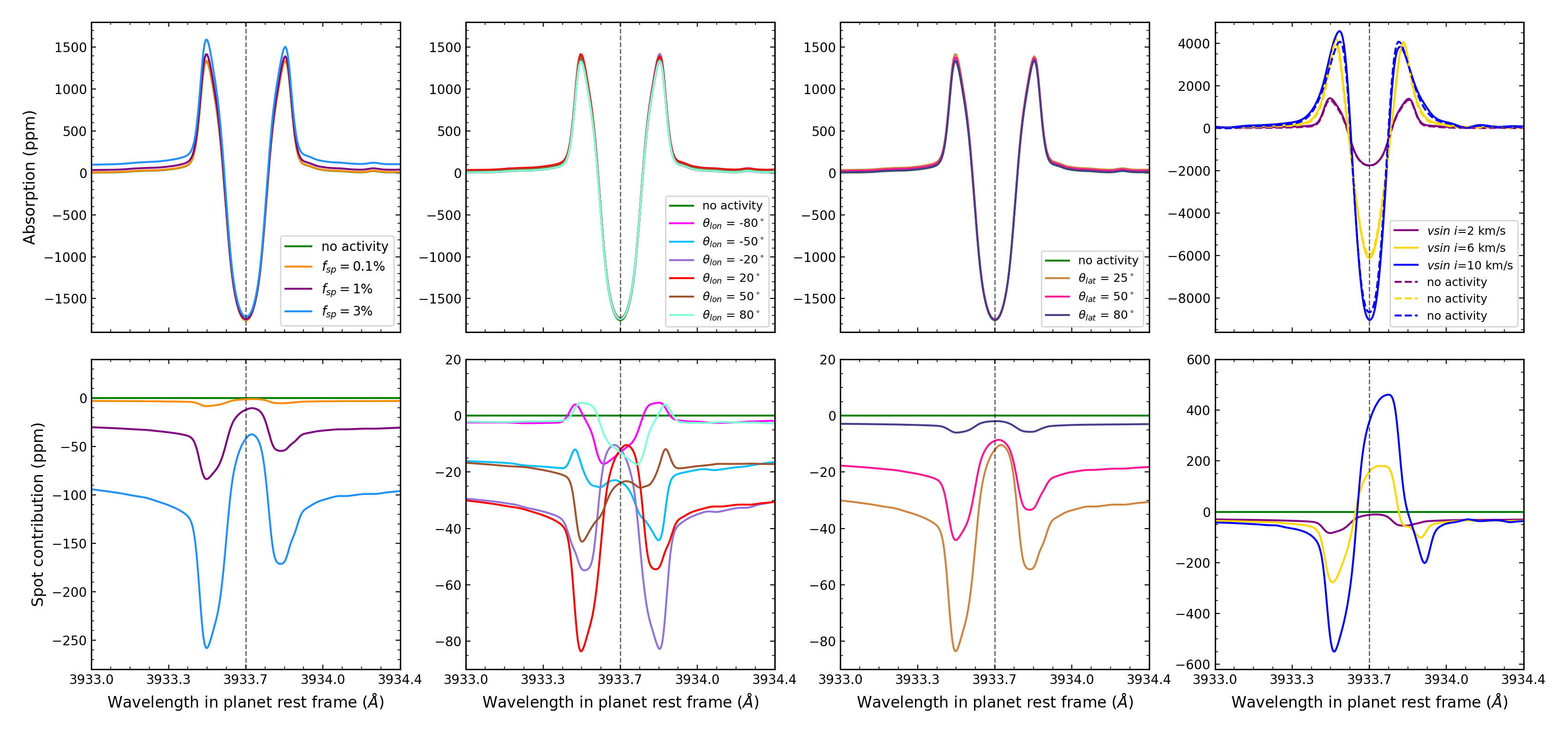}
    \caption{Top: Mean in-transit absorption spectra in the planetary rest frame around the \ion{Ca}{II} K line. Bottom: Spot contribution as the difference obtained by subtracting the quiet-star profiles from the star with a spot profiles. Each column represents (from left to right) the impact of spot size, spot position (longitude and latitude, respectively), and stellar rotational velocity. According to each case the fixed parameters were \( f_{\text{sp}} \) = 1\%, $\theta_{lat}$ = $25^\circ$, $\theta_{lon}$ = $20^\circ$ and \(v \sin i\) = $2\,\mathrm{km\,s^{-1}}$.}
    \label{fig6}
\end{center}
\end{figure*}

The same wavelength range was adopted for all three spectral lines. For H$\alpha$, previous transmission spectroscopy studies (e.g., \citealt{yuri}) have shown that the planetary absorption signal is broader than for other lines; therefore, using this range ensures that the expected planetary signal is fully encompassed in all cases. See the Appendix for a complementary analysis in which the absorption was measured using the mean around the line center. Second, we introduced a measure of asymmetry to evaluate the impact of stellar spots on the line centroid shift. 
For this purpose, we computed the spectral center of mass $\lambda_{CM}$, defined as a weighted average of the total flux:
\begin{equation} 
\lambda_{\text{CM}} = \frac{\sum_i \lambda_i F_i}{\sum_i F_i} 
\end{equation}
where \(\lambda_i\) represents the wavelengths within the line region, and \(F_i\) is the corresponding normalized flux at each \(\lambda_i\). Using the spectral center of mass $\lambda_{CM}$ we computed the velocity shift $v_{\text{shift}}$ using the non-relativistic Doppler approximation:
\begin{equation}
    v_{\text{shift}} = c \frac{\lambda_{\text{CM}} - \lambda_0}{\lambda_0}
\end{equation}
where \( c \) is the speed of light and \( \lambda_0 \) corresponds to the central wavelength of the rest-frame of each spectral line (e.g., 6562.85\, \AA \ for H$\alpha$, 5895.92\, \AA \ for \ion{Na}{I} D1 and 3933.66\, \AA \ for \ion{Ca}{II} K).

To focus on the spectral interval where the core planetary spectral line is expected, we compute both the amplitude and the line centroid over a wavelength range corresponding to $\pm 10$ $\mathrm{km\,s^{-1}}$ around the expected planetary line position. This focus ensures that the derived metrics are not affected by variations outside the location of the expected atmospheric signature. These metrics allow us to track how the line profile changes both in shape and position, providing insight into the impact of stellar contamination on velocity-resolved atmospheric signals.
Since our model is based on 1D PHOENIX atmospheres, the suppression of convective blueshift in magnetically active regions is not included. Although the current version of {\tt SOAPv4} allows such effects to be mimicked through a simplified model, we do not consider them in this work. In G-K stars, the net convective blueshift is typically of the order 400-600 $\mathrm{m\,s^{-1}}$ \citep{meunier}. A first-order scaling with the spot filling factors considered in this work (up to a few percent) indicates that neglecting this effect may lead to unmodelled velocity shifts of a few to a few tens of $\mathrm{m\,s^{-1}}$.

\section{Results of simulations}
\label{sec3}
To analyze the effect of the spot on the transmission spectra, we considered the different parameters detailed in Table~\ref{t2}. We initially focused on the \ion{Na}{I} doublet region, which is particularly sensitive to stellar activity and has been widely used in exoplanet atmosphere studies \citep[e.g.,][]{Casasayas_Barris_2019, cheng, bibiana, zhang,seidel2023, yuri}. In Figure \ref{fig4} we show the tomography maps around the Na\,\textsc{i} doublet for the simulated hot Jupiter, including the case of a stellar spot with a coverage fraction of \( f_{\mathrm{sp}} = 1\% \). The comparison between the inactive (top) and active (middle) cases reveals additional absorption during transit, when stellar activity is present. The bottom panel isolates this effect, showing that the spot contribution introduces asymmetric residuals. This illustrates that even low spot coverage can produce distortions in the transmission spectra.

\subsection{Parameter-by-parameter analysis}
\label{sec3.1}

To analyze the distortions in the line profiles caused by non-occulted stellar spots, we first examined the effect of each of the four parameters independently.
Figures~\ref{fig5}, \ref{fig6}, and \ref{fig7} show the resulting absorption spectra for the \ion{Na}{I} D1 line, \ion{Ca}{II} K, and H$\alpha$, respectively. Each column corresponds to a different parameter variation: spot filling factor ($f_{\mathrm{sp}}$), spot longitude ($\theta_{\mathrm{long}}$), spot latitude ($\theta_{\mathrm{lat}}$), and projected stellar rotational velocity ($v \sin i$), from left to right. The top panels display the absorption spectra \((1 - F_{\text{in}} / F_{\text{out}})\) as a function of wavelength in the planetary rest frame, while the bottom panels show the difference between the spotted and unspotted spectra, effectively isolating the spot contribution. As illustrated, both the amplitude and shape of the absorption signal can be significantly altered by stellar activity, depending on the specific spot configuration and stellar rotation. 

In this context, we adopt the term POLDs, following \citet{Dethier_2023}, to refer to the distortions originating from the non-similarity between the locally occulted stellar line and the disk integrated stellar lines. In our usage, POLDs will trace the impacts of stellar rotation, limb-darkening, and the stellar spot, but exclude the impact of line profile variations as a function of position on the disk.
\subsubsection{Effect of spot size}
\begin{figure*}[h]
\begin{center}
\includegraphics[scale=0.43]{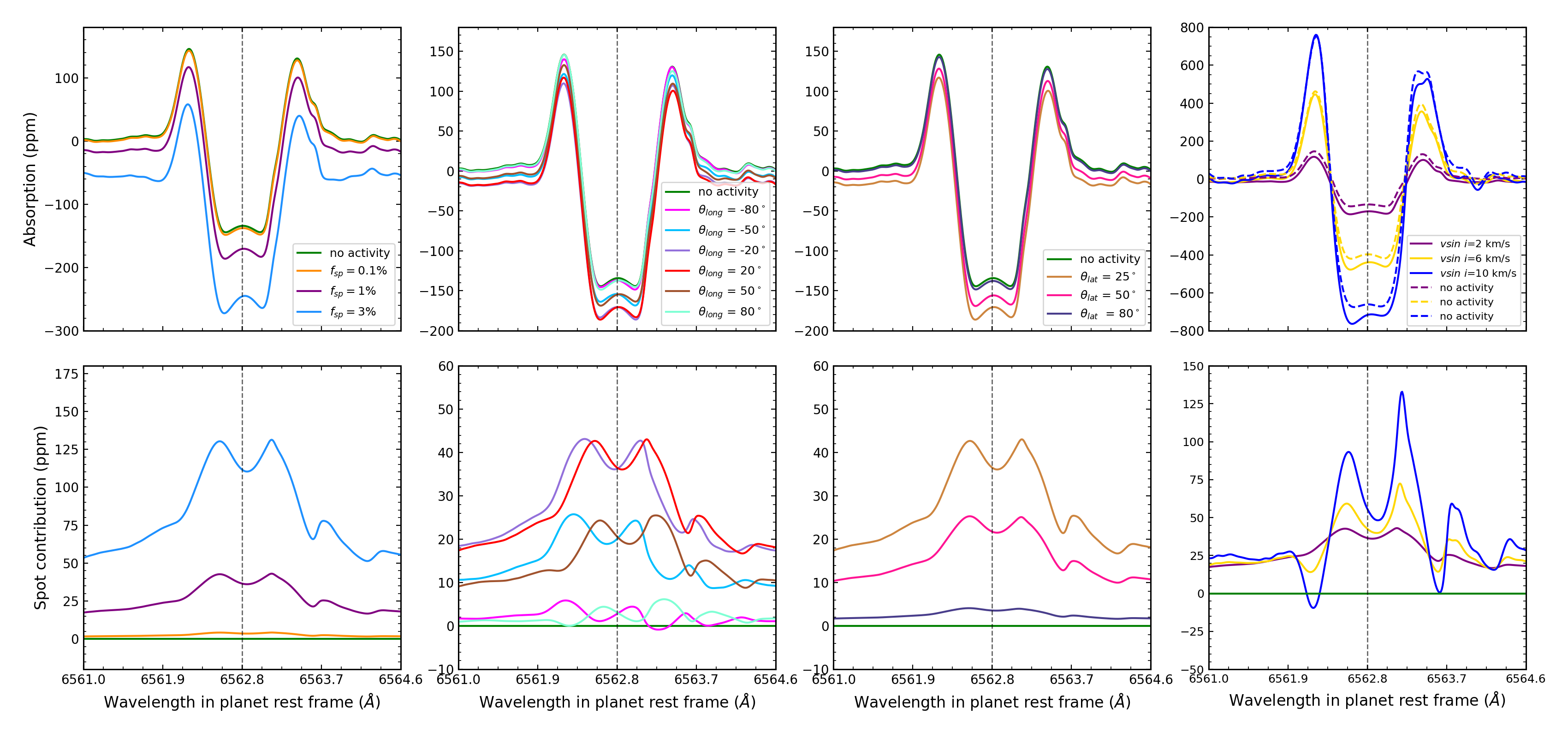}
    \caption{Top: Mean in-transit absorption spectra in the planetary rest frame around the H$\alpha$ line. Bottom: Spot contribution as the difference obtained by subtracting the quiet-star profiles from the star with a spot profiles. Each column represents (from left to right) the impact of spot size, spot position (longitude and latitude, respectively), and stellar rotational velocity. According to each case the fixed parameters were \( f_{\text{sp}} \) = 1\%, $\theta_{lat}$ = $25^\circ$, $\theta_{lon}$ = $20^\circ$ and \(v \sin i\) = $2\,\mathrm{km\,s^{-1}}$.}
    \label{fig7}
\end{center}
\end{figure*}

In the first column of Figures \ref{fig5}, \ref{fig6}, and \ref{fig7}, we show the variation of the distortion as a function of spot sizes for each line We observe that the contribution of the spot increases with its coverage fraction \( f_{\text{sp}} \), as expected. Larger spots introduce more significant distortions in the absorption profile for the three lines, which is consistent with previous studies on stellar activity contamination in transmission spectra \citep{pont, Rackham_2018}. However, those works focused on low-resolution spectra, where spot-induced distortions manifest primarily as wavelength-independent offsets. In contrast, the high-resolution framework adopted here reveals localized anomalies in the line profile, allowing us to directly resolve the spectral imprint of stellar heterogeneities.

The line \ion{Ca}{II} K exhibits the strongest response to increasing \( f_{\text{sp}} \), with amplitudes (measured from peak to peak) exceeding 250 ppm even for the highest activity case we explored (3\% of $f_{\text{sp}}$). The H$\alpha$ line, on the contrary, shows a more modest increase varying up to 125 ppm. In the H$\alpha$ region, the local continuum is significantly affected by the spot. In Figures \ref{fig2} and \ref{fig3}, we can observe how the spot spectrum lies above the quiet photosphere spectrum. This contrast arises from the different formation physics of the lines \citep{Gray_2005}. Interestingly, the H$\alpha$ POLDs show a small bump around the center of the line that is not present in the other lines. The origin of this characteristic is not entirely clear, but it could be related to the larger intrinsic width of H$\alpha$ and the way the transmission spectrum is computed. The averaging process over orbital phases during the transmission spectrum extraction may enhance small residuals in the line core, producing the observed bump.
In Figure \ref{fig15} we show the tomography plot around the H$\alpha$ line to identify how the POLD behaves in the closest lines to H$\alpha$ to also understand the origin of the bump. We also show the results in Figure \ref{fig16} for \ion{Na}{i}~D1 for comparison.  The comparison reveals that this feature is unique to H$\alpha$, while its exact origin remains uncertain. 

\subsubsection{Effect of stellar rotation}
It has already been established that increasing \( v \sin i \) enhances the amplitude of line-profile distortions \citep{Cegla_2016, Dethier_2023}. In our simulations, we further show that, when starspots are present, the amplitude of spot-induced contribution to the POLDs increases approximately proportionally with \( v \sin i \), while also affecting the detailed morphology of the features. As \(v \sin i\) increases from 2 to 10~km\,s\(^{-1}\), the amplitude variations caused by unocculted spots become increasingly pronounced, with amplitudes exceeding 500~ppm for \ion{Na}{I} D1. 
This effect is particularly evident in the \ion{Ca}{II} K line (see Figure \ref{fig6} fourth column bottom panel), where distortions can exceed 1000~ppm at the highest rotation rates tested. The observed distortions result from a combination of factors, including the local modification of the line profile by the spot, the broadening of the disk-integrated line due to stellar rotation, and the motion of the spot across the stellar disk during transit. The interplay of these effects makes it difficult to disentangle and quantify the exact contribution of each factor to the amplitude of the distortions.
\subsubsection{Effect of spot position}
For the spot longitude, we find that the contribution is symmetric with respect to the central wavelength of the lines for each pair of \( \pm \theta_{\text{lon}} \) and it remains within the same order of magnitude. This behavior results from the combination of limb-darkening and local Doppler shifts, which modulate the flux contribution of each surface element as the spot moves across the stellar disk. However, for H$\alpha$ we observe that the spot contribution is not entirely symmetric. This effect might be explained by the presence of the secondary component on the right-hand wing of the H$\alpha$ line, which is quite pronounced in the spot spectrum.
For spot latitudes, we observe that spots at lower latitudes contribute more strongly to the absorption profile distortions for the three spectral lines. This effect could be attributed to limb-darkening, as lower-latitude spots tend to be projected against brighter regions of the stellar disk, enhancing their impact on the integrated spectra. 
This is more evident in the \ion{Ca}{II} K and \ion{Na}{I} D1 lines, where the flux contrast is higher near the disk center. From our simulations, we observed that spots at higher latitudes have a weaker impact because both their flux contrast and their projected area in the observer’s line of sight are reduced.
\begin{figure*}
\centering
\includegraphics[scale=0.44]{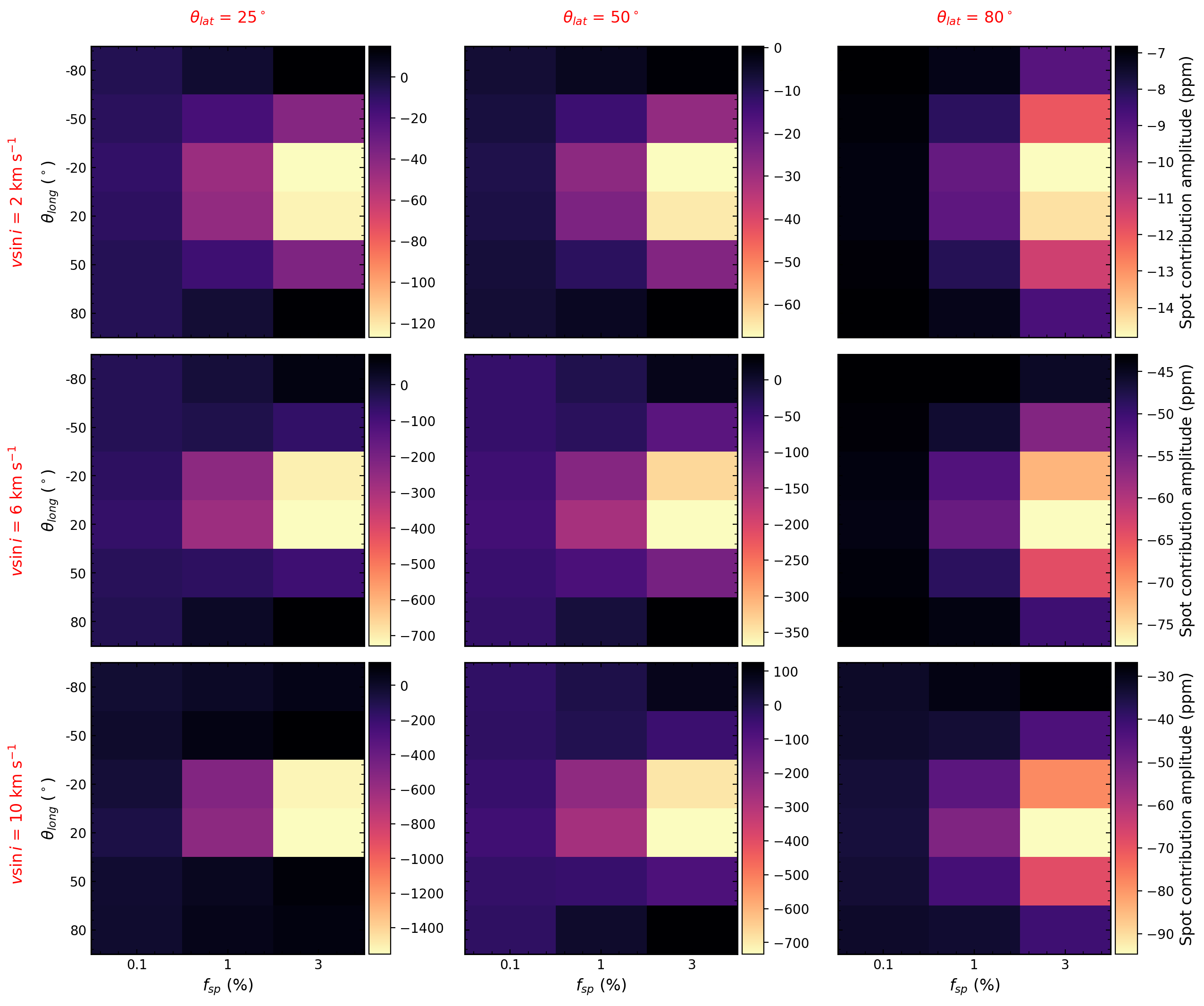}
\includegraphics[scale=0.42]{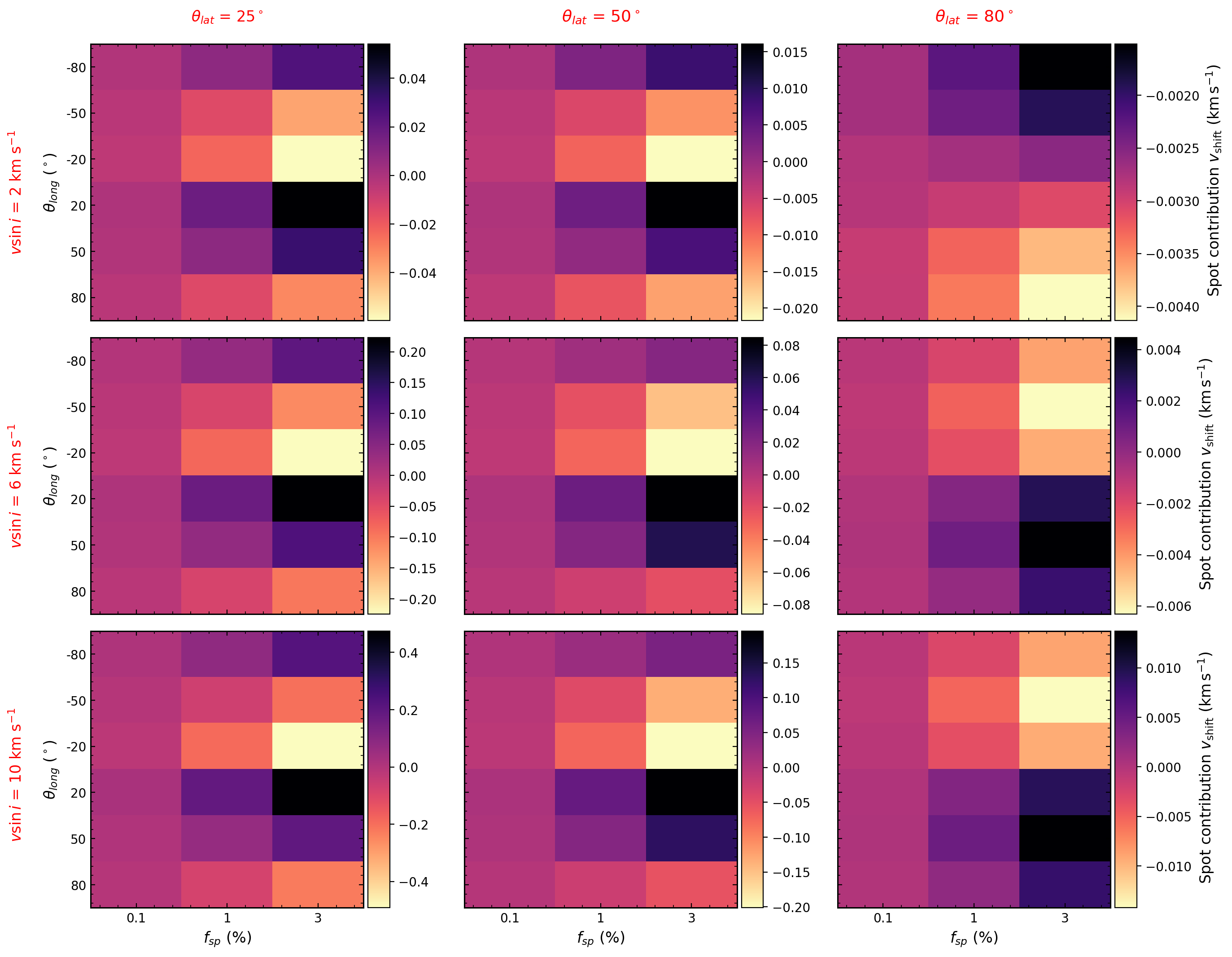}
\caption{2D maps of the amplitude (top) and velocity shift (bottom) of the absorption spectrum of the \ion{Na}{I} D$_1$ line.}\label{fig8}
\end{figure*}
\subsection{Grid of combined parameters}
\label{sec3.2}
After allowing all four parameters to vary, we conducted 216 simulations covering different combinations of \( v \sin i \), \( \theta_{\text{lat}} \), \( \theta_{\text{lon}} \), and \( f_{\text{sp}} \). The results are presented in Figures~\ref{fig8}, ~\ref{figA1} and \ref{figA2}, which show how the amplitude and velocity shift of the spot contribution change as a function of spot size, longitude, \( v \sin i \) and latitude for the different spectral regions. \\ \\
In all cases, we observe that the spot contribution decreases as the active region moves toward the stellar limb, consistent with the reduced projected area and the partial coverage of the spot on the far hemisphere. The apparent abrupt variations seen in some configurations correspond to situations where part of the spot becomes partially hidden, rather than to intrinsic changes in the line depth. Although increasing \( f_{\text{sp}} \) still enhances the amplitude of the spot contribution in all lines, rotation has a stronger impact on the amplitude of the lines, since it broadens the stellar spectrum \citep[e.g.,][]{Cegla_2016}.
We observe that for each pair of $\pm\theta_{\mathrm{lon}}$, the absorption values remain similar. This symmetry is due to the fact that the spots occupy symmetric positions in velocity across the stellar surface, combined with an equatorial transit. This symmetry could break if any of these conditions changes. For H$\alpha$, the deviations from the exact symmetry likely arise from the presence of a very narrow line near the core, which is probably the culprit. 
In our 2D analysis, we find no strong relationships between the spot contribution amplitude and individual parameters such as spot latitude, longitude, filling factor, or stellar $v \sin i$, except for a clear link between spot longitude and latitude. This link likely arises from the dependence of the projected spot area on the spot’s position on the stellar disk, which remains the dominant factor controlling the amplitude.\\ \\
In addition to the changes in the amplitude, we also examined the velocity shifts of the lines induced by stellar spots. We find that for low rotational velocities (\( v \sin i = 2 \) km s\(^{-1}\)), the induced changes are smaller and spatially variable. However, in all configurations except for $\theta_{lat}$ = $80^\circ$ a clear and consistent pattern emerges: the velocity shift maps become morphologically similar, differing only in scale. The maximum shifts occur when the spot is near the center of the stellar disk (e.g., $\theta_{\rm lat} \sim \pm 25^\circ$), where its contribution to the integrated stellar spectrum is largest. Moreover, the shift increases with spot size, consistent with a larger spot covering more of the stellar surface and thus having a greater effect on the centroid displacement.

We also observe that, in terms of asymmetry, H$\alpha$ is the most affected line, showing shifts in spot contribution from -10 $\mathrm{km\,s^{-1}}$ to 6 $\mathrm{km\,s^{-1}}$, while \ion{Na}{I} D1 and \ion{Ca}{II} K exhibit much smaller shifts, ranging from -0.4 $\mathrm{km\,s^{-1}}$ to 0.4 $\mathrm{km\,s^{-1}}$. The larger velocity range observed for H$\alpha$ can be explained by its broader line core, which allows spot-induced perturbations to contribute over a wider wavelength interval, effectively amplifying the apparent displacement of the line centroid. In contrast, the narrower cores of \ion{Na}{I} D1 and \ion{Ca}{II} K restrict these perturbations to a smaller region, leading to smaller velocity shifts.

\section{Observational relevance}
\label{sec4}
To assess the observational relevance of the spot-induced signal, we compare our results with Na\,\textsc{i} transmission measurements reported in the literature for well-studied hot Jupiters. We chose to focus here on 
Na\,\textsc{i} because is the signature that is most commonly detected in hot jupiters, but a similar comparison could be made for Ca\,\textsc{ii} and H$\alpha$ \citep[e.g.,][]{Cauley_2018}. 
Our simulations explore a specific and configuration, consisting of a single non-occulted active region, with some changes in some of the parameters. Other stellar surface configurations are not considered in this work.

For WASP-127b, ground-based high-resolution observations report integrated Na\,\textsc{i} absorption depths of approximately $3400 \pm 400$~ppm, \citep{allart}, while for WASP-52b yield a broadband Na\,\textsc{i} absorption of $10900 \pm 1600$~ppm \citep{cheng}.
For the more active system HD~189733~b, high-resolution spectroscopy reveals stronger Na\,\textsc{i} absorption, with reported depths ranging from roughly $2000$ to $7000$~ppm \citep{khalafinejad}, and typical statistical uncertainties of several tens to a few hundred ppm, depending on the adopted passband.
In our simulations, except for the high \(v \sin i\) and high-activity cases, the isolated effect of unocculted spots produces distortions at the level of $10$--$100$~ppm across the Na\,\textsc{i} line. Although this contribution is smaller than the planetary sodium signal itself, its amplitude is comparable to the typical observational uncertainties reported for high-resolution Na\,\textsc{i} measurements, from 79 ppm up to 600 ppm  (e.g.\ \citealt{Azevedo_Silva_2022, Casasayas_Barris_2017}). Unocculted spots can therefore bias the inferred Na\,\textsc{i} absorption depth at the level of current observational precision if not properly accounted for. 

On the other hand, high-resolution transmission spectroscopy has also been used to infer atmospheric dynamics from resolved sodium line profiles, including efforts to retrieve wind velocities from the Na \textsc{i} doublet (e.g. \citealt{Seidel2020}). General circulation models predict day-to-night winds in hot-Jupiter atmospheres that can produce net blueshifts of the Na \textsc{i} transmission signal at the level of several km s$^{-1}$ (e.g. \citealt{Louden_2015}). Such blueshifts have indeed been reported in high-resolution observations of HD~189733b (e.g. \citealt{Wyttenbach, mounzer}).

In our simulations, although the velocity shift metric is not directly comparable to atmospheric wind velocities inferred from transmission spectra, its order of magnitude provides a useful benchmark to assess the potential impact of stellar activity on dynamical interpretations. For Na \textsc{i} D lines, spot-induced distortions lead to apparent centroid displacements in the range of $-0.4$ to $+0.4$~km s$^{-1}$. At this level, stellar activity alone cannot account for km s$^{-1}$ wind measurements. However, velocity shift values approaching the km s$^{-1}$ regime would indicate that stellar activity could significantly bias the interpretation of atmospheric dynamics and should be investigated in more detail.
\section{Parameter Impact Analysis}
\label{sec5}
\subsection{Effect of the spot-induced RV slopes}
\label{sec4.1}
\begin{figure}[h]
\begin{center}
\includegraphics[scale=0.35]{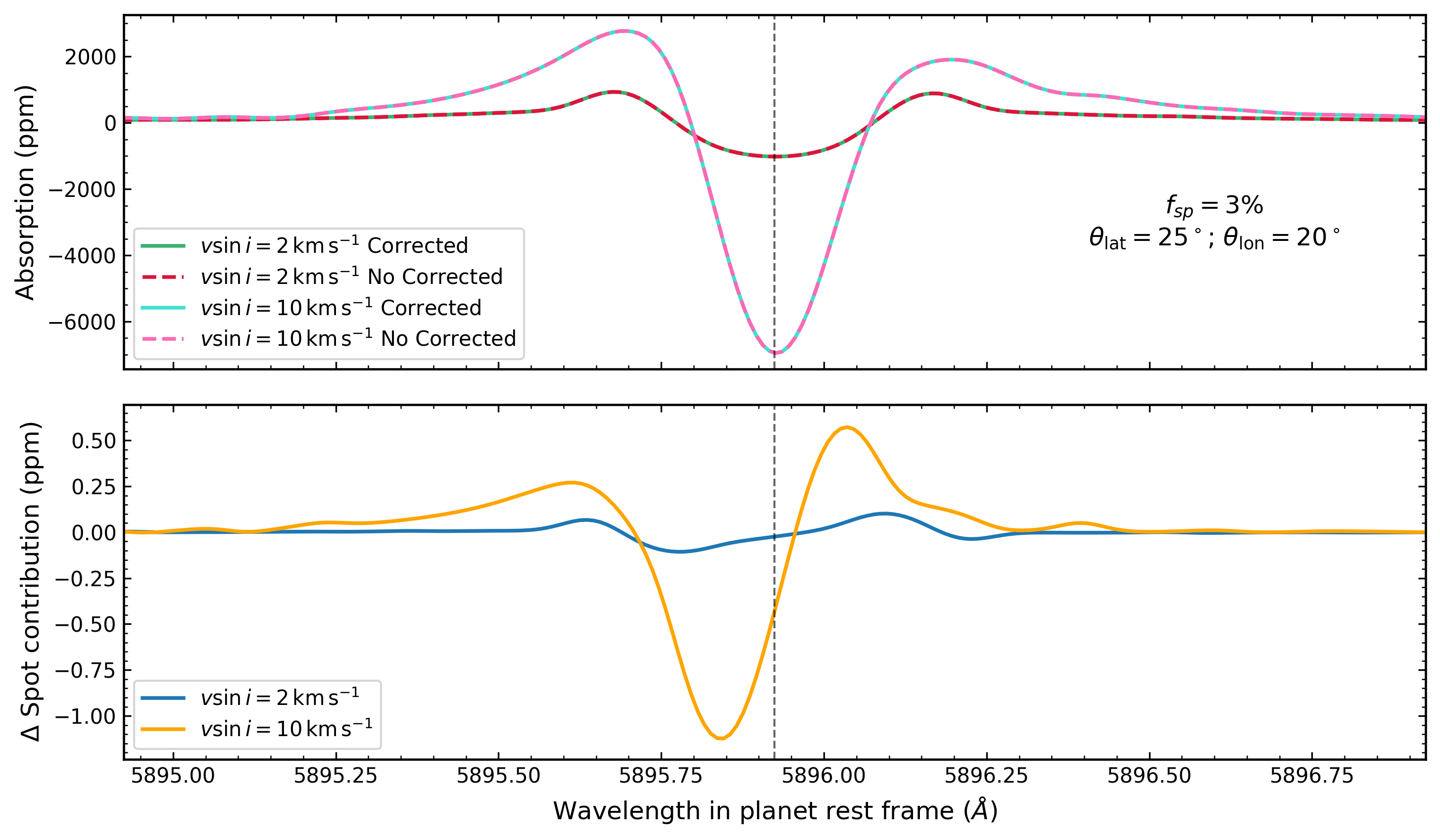}
\caption{Comparison between absorption spectra with and without the correction for the spot velocity, illustrated for the \ion{Na}{I} D$_1$ line and for two different \(v \sin i\) values. Top: shows the averaged absorption spectra for each case: in green and cyan, the POLD shifted by the spot-induced velocity; in red and pink dashed, the POLD computed using only the Keplerian velocity of the planet. Bottom Panel: shows the relative difference contribution of the spot for each of the cases.}
\label{fig11}
\end{center}
\end{figure}
In addition to their direct impact on the planetary transmission spectrum, stellar spots are known to affect the measured radial velocities (RVs) of the star. During a typical transmission spectroscopy observation lasting several hours, spots introduce both a shift and a slope in the stellar RV time series, which can bias our estimates of the systemic velocity and the star's Keplerian motion. These quantities are used to shift the spectra into the stellar and planetary rest frames, meaning that spots can deform stellar and planetary lines purely through their influence on the RV measurements.

To assess the impact of the spot-slope effect on the transmission spectrum, we compared two sets of simulations: one using the exact systemic velocity and Keplerian motion of the star and planet, and another measuring these quantities from the RV time series affected by the spots and propagating the impact on the slope when moving to the stellar rest-frame. We are not propagating the impact of the slope on the systemic velocity as it would only produce an overall shift of the planetary lines.
Figure~\ref{fig11} illustrates this comparison for the \ion{Na}{I} D1 line, where differences between the spot contribution for both cases and considering our highest activity case $f_{sp}$ of 3\% reach up to $\sim$0.2~ppm for a $v \sin i$ of 2 $\mathrm{km\,s^{-1}}$, and up to $\sim$1.5~ppm for a $v \sin i$ of 10 $\mathrm{km\,s^{-1}}$ in specific regions of the line profile, particularly near the core and wings.

For the case of a hot Jupiter and for our highest $f_{sp}$ value considered, even considering the case of a fast rotator star ($v \sin i$ = 10 $\mathrm{km\,s^{-1}}$), this difference represents a small fraction of the total transmission signal and does not significantly affect the inferred atmospheric properties. This indicates that neglecting the impact of the spot on the measured RV semi-amplitude, as commonly done in current analyzes of giant exoplanets, does not introduce significant biases in the interpretation of giant exoplanet atmospheres. Upcoming facilities such as the ELT and PLATO, with unprecedented precision, will target smaller exoplanets, for which this effect might be more relevant and should be assessed. 
\subsection{Impact of spectral normalization}
\begin{figure}[h]
\begin{center}
\includegraphics[scale=0.4]{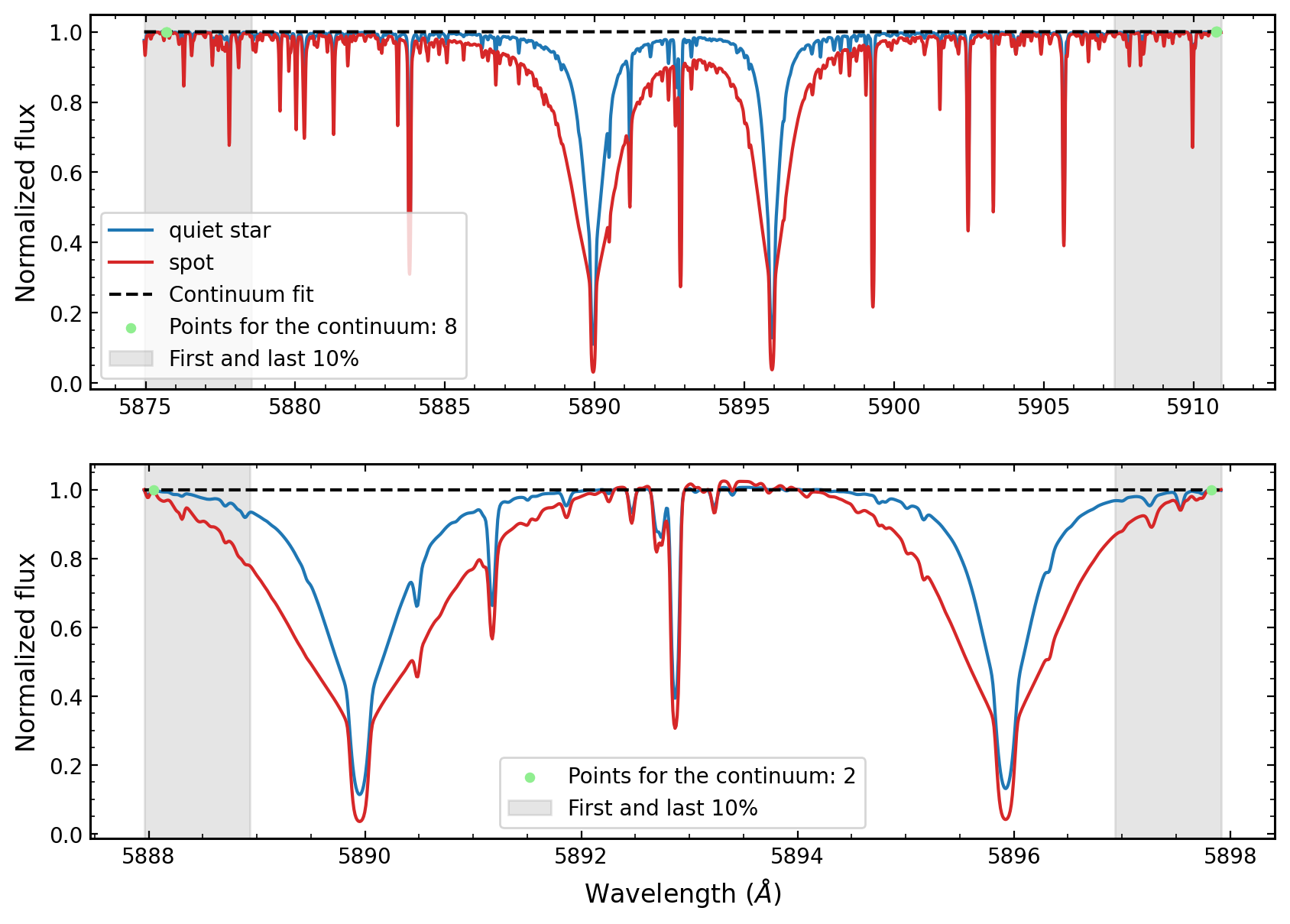}
\caption{\ion{Na}{I} D spectral region. Top: wide normalization range, using the first and last 10\% of the total spectral window (gray regions), with 8 points used to estimate the continuum (green dots). Bottom: narrow window, where only 2 continuum points are available, resulting in a poorer continuum estimate.}
\label{fig122}
\end{center}
\end{figure}

In high-resolution transmission spectroscopy, normalization accuracy critically depends on how the stellar continuum is defined. Figure~\ref{fig122} illustrates that the choice of the normalization window strongly influences the estimated continuum level. In the \ion{Na}{i} doublet region, using a narrower wavelength range (bottom panel) results in a poorer continuum estimate due to the lack of truly line-free regions. Only a few points can be used for the continuum fit, and they are partially affected by the line wings, leading to an underestimation of the baseline flux. Since the lines are broader in the spot, this effect is more pronounced in the spot spectrum than in the quiet Sun spectrum. Consequently, the spot line contrast is overestimated relative to that of the quiet Sun line. 
In contrast, using a wider spectral window (top panel) expands the region available for the normalization algorithm to identify continuum points. Since the search is restricted to this window, its width directly limits which regions can be sampled as a continuum. In the line wings, only a few points fall within the window on each side, so the normalization reliability ultimately depends on the chosen window width.

To assess how the choice of normalization window affects the resulting spectral metrics, we repeated the full set of 216 simulations described in Section \ref{sec3.2} using a narrower wavelength range (see Figure \ref{fig122}, bottom panel). We found that both the measured absorption and asymmetry values differ between the two configurations. When considering the isolated contribution of the unocculted spot (i.e. after removing the RME and CLV components), the absorption amplitude can vary by up to 175\% in relative terms, while the line asymmetry varies by up to 200\%, depending solely on the normalization window used.

This analysis highlights that normalization errors can arise not from the fitting method itself but from suboptimal choices of the spectral region used for fitting. Although in our simulations a broader fitting window is easy to achieve, several factors can in practice degrade the accuracy of the continuum estimation: broad lines, nearby telluric features, crowded spectral regions, and the borders of spectral orders. The narrower windows tested here for \ion{Na}{I} thus illustrate how poorly defined continuum regions can lead to biases.

\section{Conclusions}
In this work, we present results from a diagnostic investigation of the impact of stellar spots on high-resolution transmission spectra. By simulating planetary transit spectra with the {\tt SOAPv4.0} code, we quantified how spots modify the shape and amplitude of the POLDs in different spectral regions of the absorption spectrum of a transiting hot Jupiter. We examine a broad parameter space covering stellar rotation rates, spot sizes, and positions for a hot Jupiter system. Our investigation demonstrates that the distortions introduced by non-occulted stellar spots in high-resolution transmission spectra arise from a complex interplay of stellar activity and rotation.

When we only vary one parameter, the spot coverage fraction ($f_{\mathrm{sp}}$) remains the primary driver of the distortion amplitude. The line \ion{Ca}{II} K shows the strongest response, with spot contribution amplitudes above 250 ppm for our highest $f_{\mathrm{sp}}$ value tested and for the slow rotator case, while \ion{Na}{I} D1 displays intermediate values. In contrast, H$\alpha$ exhibits weaker but still distinctive distortions, including a characteristic central bump, which may be associated with the relatively broad core of the line compared to the others.

Increasing $v \sin i$ further amplifies both the amplitude and the complexity of the signatures, with distortions exceeding 2000 ppm in \ion{Ca}{II} K at the highest rotation rates. The combined effects of local spot modifications, rotational broadening, and spot motion make the resulting features highly structured and variable.

The spot geometry also contributes in systematic ways. Low-latitude spots have the strongest impact due to their projection against brighter disk regions, while longitude introduces nearly symmetric contributions for $\pm \theta_{\mathrm{lon}}$.
When multiple parameters vary simultaneously, the dominant factor controlling the distortion amplitude is the projected spot area on the stellar disk. This explains the clear relationship between spot longitude and latitude found in our 2D grid simulations. 
Our simulations reveal velocity shifts in the POLDs induced by stellar spots. These shifts scale with both $v \sin i$ and the spot size, reaching their maximum when the spot lies near the center of the stellar disk. The effect is line-dependent: H$\alpha$ shows the largest excursions (from –10 to +6 $\mathrm{km\,s^{-1}}$), while \ion{Na}{I} D1 and \ion{Ca}{II} K remain confined within $\pm$0.4 $\mathrm{km\,s^{-1}}$. The broader core of H$\alpha$ distributes the spot-induced distortions across a wider wavelength range, enhancing its apparent asymmetry.

Unlike the broadband offsets reported in low-resolution studies, our high-resolution approach resolves localized line-shape anomalies and velocity shifts, underscoring the need for detailed modeling of stellar heterogeneities to disentangle planetary signals from stellar activity.

Neglecting the local velocity of the unocculted spots can introduce residual distortions of less than 2 ppm in the \ion{Na}{I} D1 line for our highest activity level and highest rotational velocity case tested.
The accuracy of the spectral normalization window also plays a critical role: narrow windows can underestimate the continuum, leading to amplitude differences of up to 175\% and relative asymmetry variations of up to 200\%.

\begin{acknowledgements}
This work was funded by the European Union (ERC, FIERCE, 101052347). Views and opinions expressed are however those of the author(s) only and do not necessarily reflect those of the European Union or the European Research Council. Neither the European Union nor the granting authority can be held responsible for them. This work was supported by FCT - Fundação para a Ciência e a Tecnologia through national funds and by FEDER through COMPETE2020 - Programa Operacional Competitividade e Internacionalização by these grants: UIDB/04434/2020; UIDP/04434/2020.
- ODSD acknowledges support from e-CHEOPS: PEA No 4000142255
- ODSD acknowledges funding by the European Union (ERC, FIERCE, 101052347)
\end{acknowledgements}


\clearpage 
\onecolumn 
\begin{appendix}
\section{Grid of combined parameters: Simulation results for \ion{Ca}{II} K and H$\alpha$}
\begin{figure*}[ht!]
\centering
\includegraphics[width=0.65\textwidth]{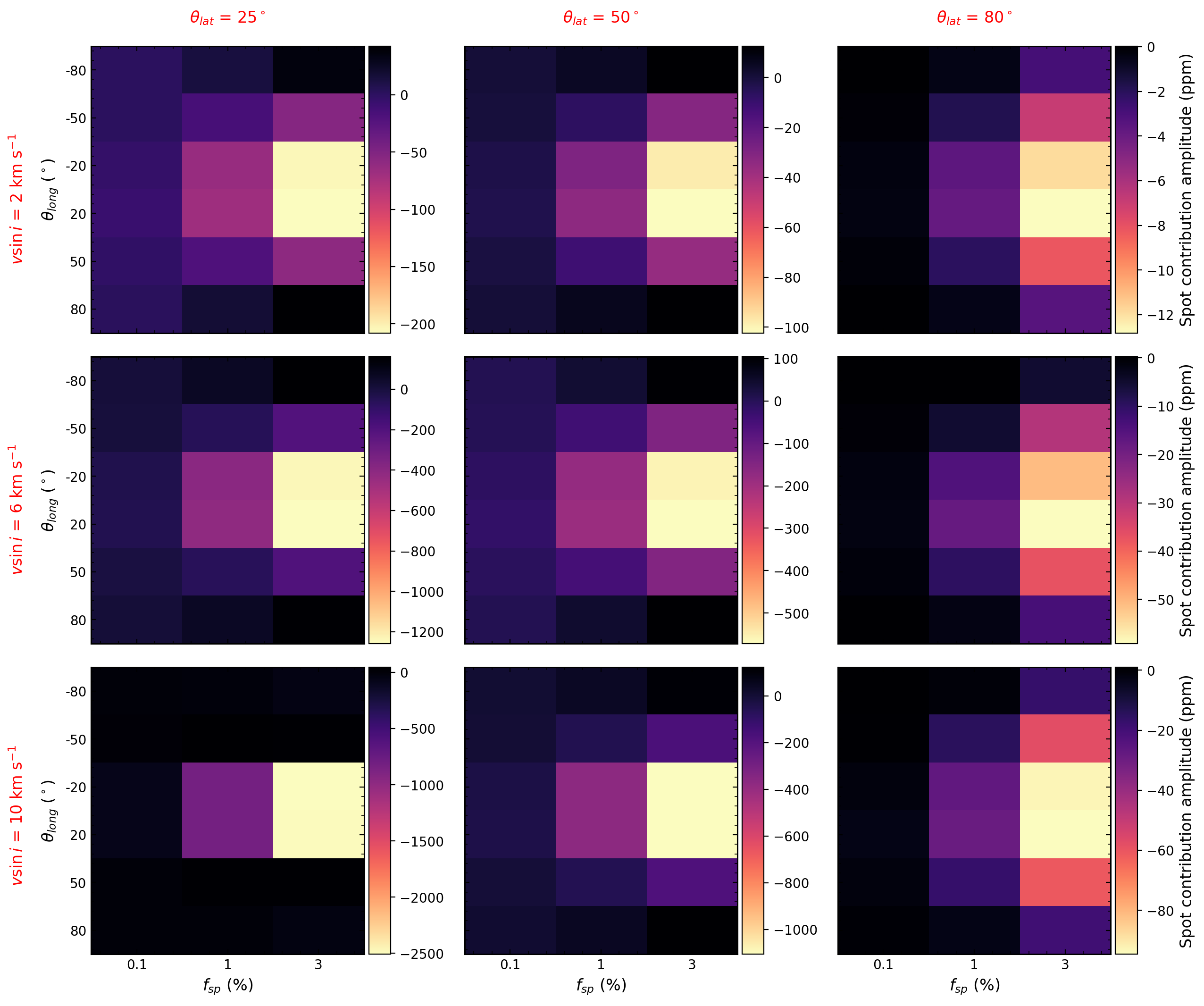}
\includegraphics[width=0.65\textwidth]{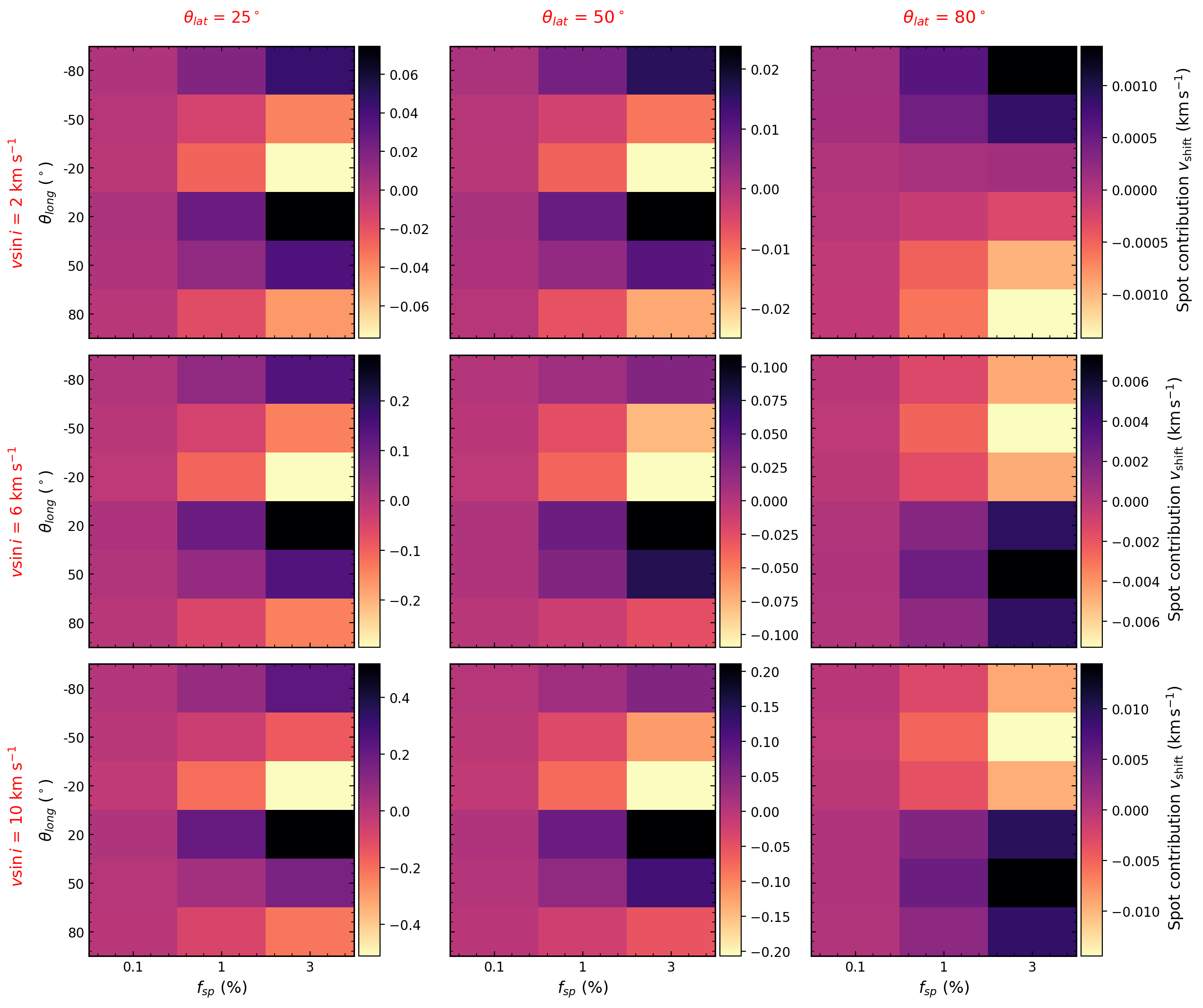}
\caption{Top: 2D absorption maps for the simulations conducted for the \ion{Ca}{II} K line. Bottom: Same as top but for asymmetry.}
\label{figA1}
\end{figure*}
\begin{figure*}
\centering
\includegraphics[width=0.72\textwidth]{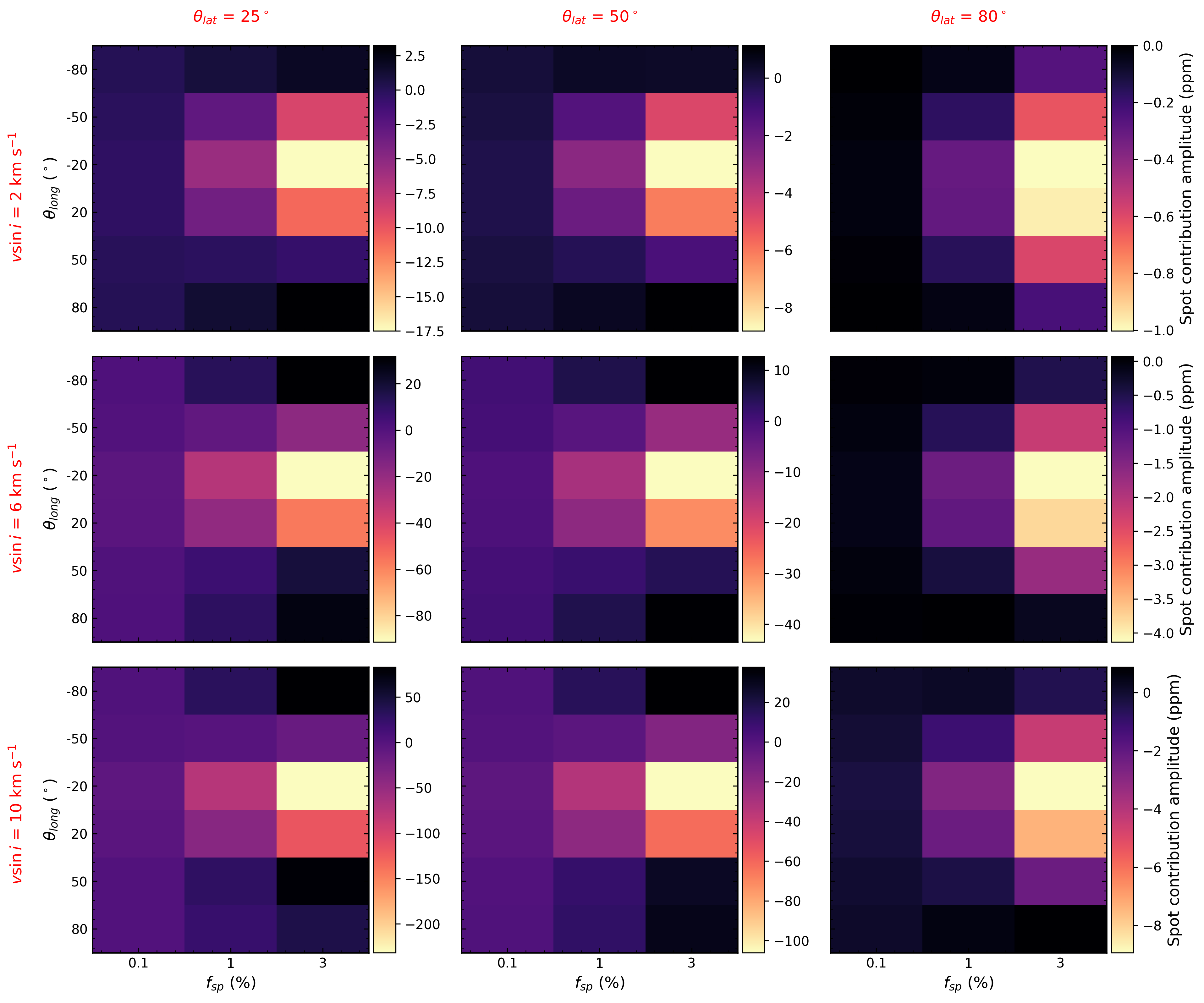}
\includegraphics[width=0.72\textwidth]{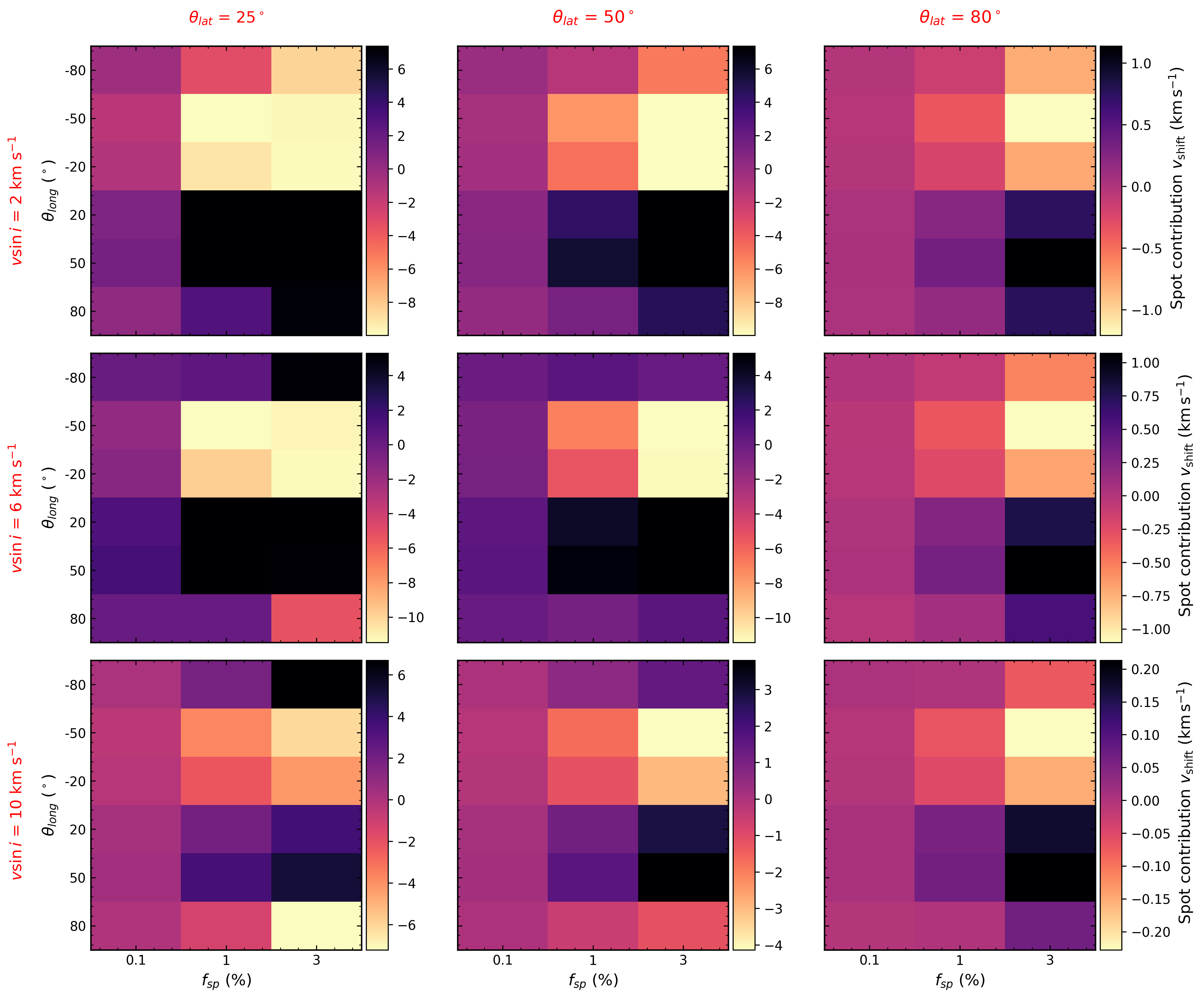}
\caption{Top: 2D absorption maps for the simulations conducted for the $H\alpha$ line. Bottom: Same as top but for asymmetry.} 
\label{figA2}
\end{figure*}
\clearpage 
\onecolumn 
\section{Alternative Approach to Absorption Profile Analysis}
\begin{figure*}[ht!]
\begin{center}
\includegraphics[width=0.7\textwidth]{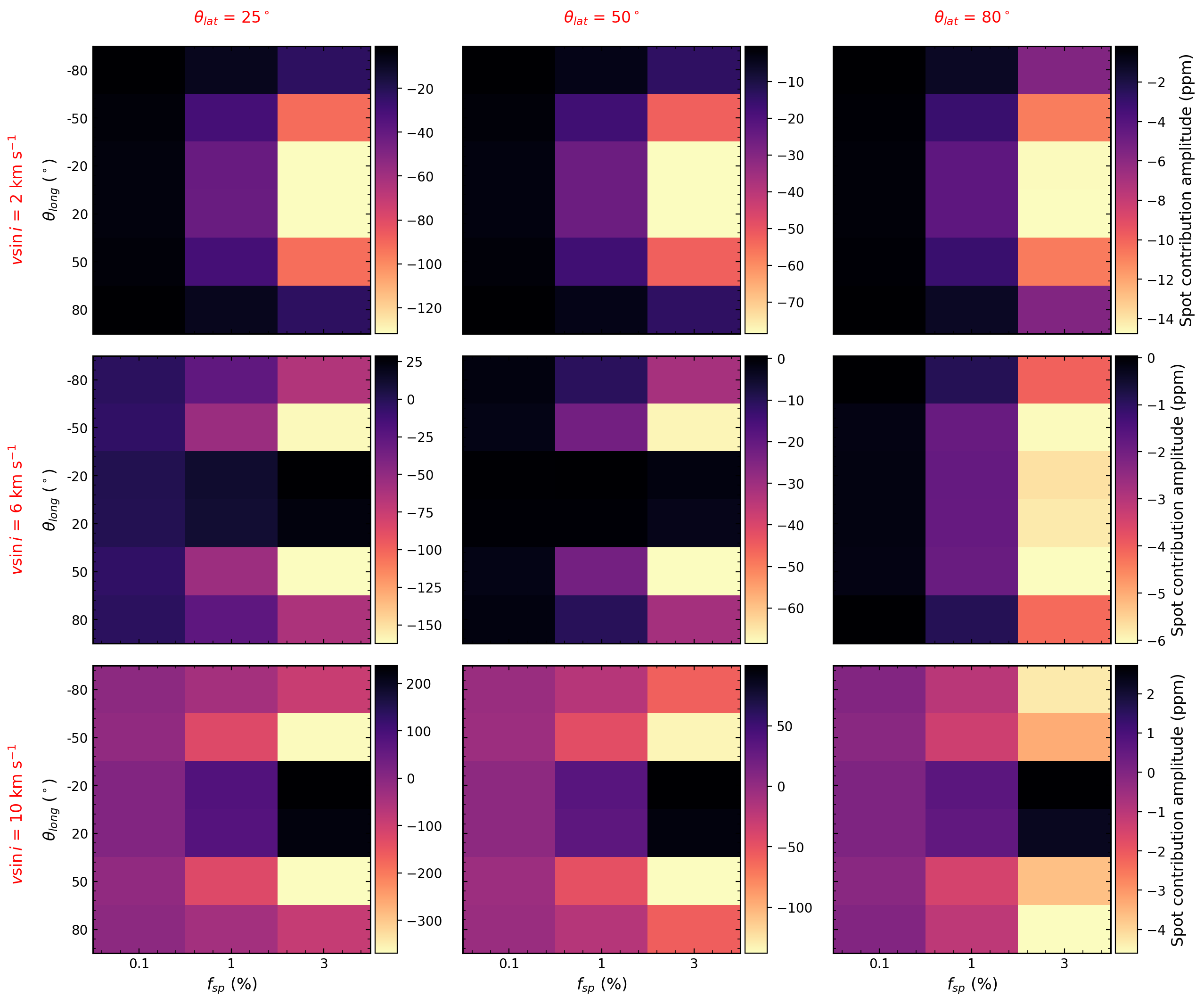}
\caption{2D absorption maps for the simulations conducted for the \ion{Na}{I} D$_1$ line}
\label{figB1}
\end{center}
\end{figure*}
In addition to using the min--max method to calculate the absorption, we also computed the mean flux around the line center. Specifically, we averaged the flux within a 10 km/s window on both the red and blue sides of the line center.
\clearpage
\section{Tomography plots around H$\alpha$ \& \ion{Na}{I}}
\begin{figure*}[ht!]
\begin{center}
\includegraphics[scale=0.50]{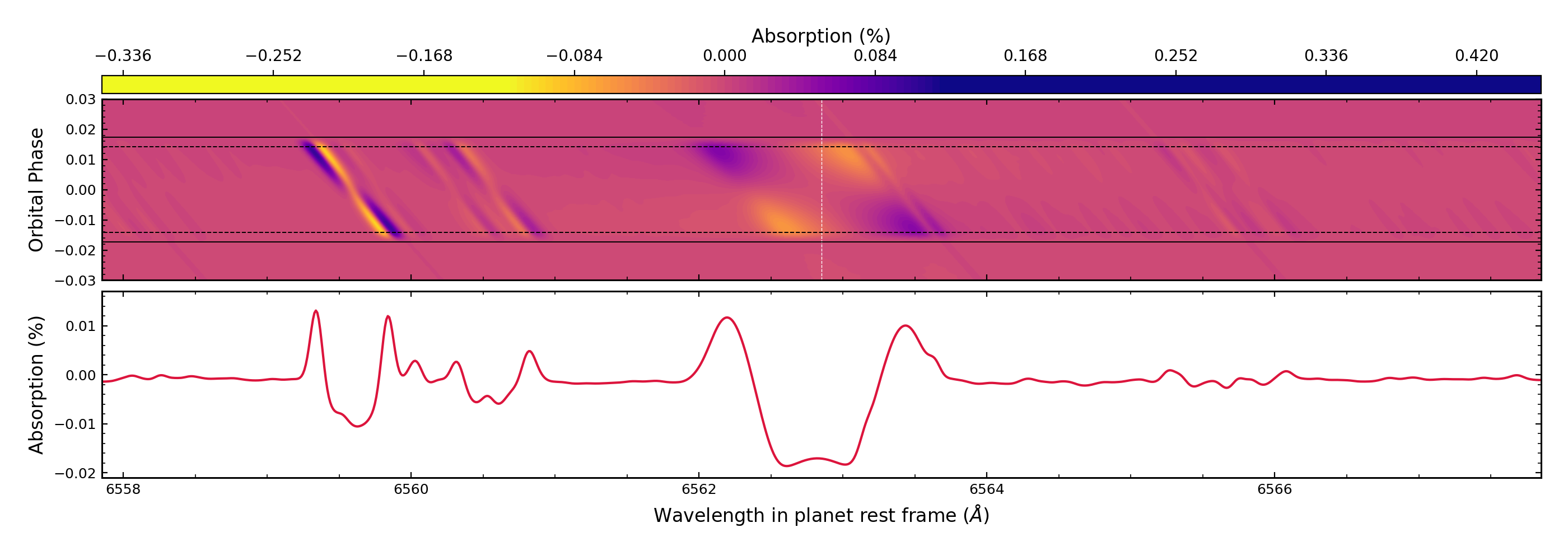}
\caption{Top Panel: Tomography plot of the individual absorption spectra near the H$\alpha$ line for a simulated hot Jupiter, including the effect of a stellar spot. Bottom Panel: Mean in-transit planetary absorption spectra around the H$\alpha$ line. The black horizontal lines represent the four transit
contacts, while the white dashed line traces the center of the line. Both plots are shown in the planet's rest frame. For this case, the spots parameters are: \( f_{\text{sp}} \) = 1\%, $\theta_{lat}$ = $25^\circ$, $\theta_{lon}$ = $20^\circ$.}
\label{fig15}
\end{center}
\end{figure*}
\begin{figure*}[ht!]
\begin{center}
\includegraphics[scale=0.515]{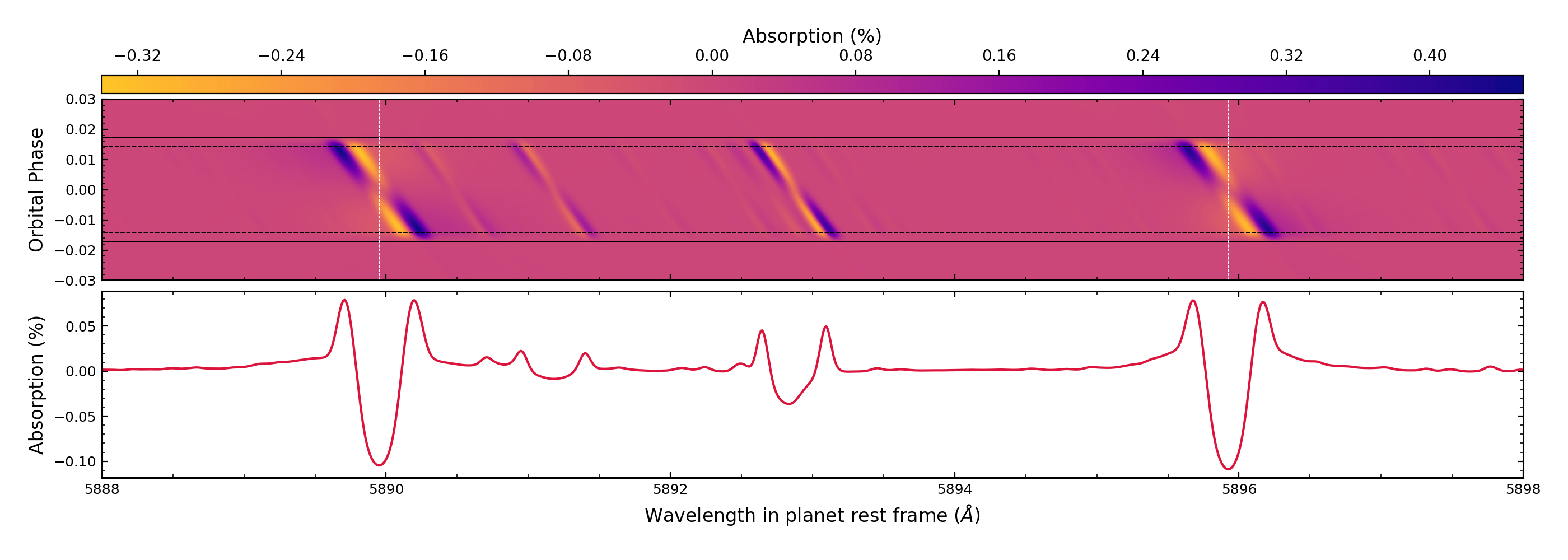}
\caption{Top Panel: Tomography plot of the individual absorption spectra near the \ion{Na}{I} doublet for a simulated hot Jupiter, including the effect of a stellar spot with (\(f_{\text{sp}}\)) of 1\%. Bottom Panel: Mean in-transit planetary absorption spectra around the \ion{Na}{I} doublet. The black horizontal lines represent the four transit
contacts, while the white dashed line traces the center of the line. Both plots are shown in the planet's rest frame. Both plots are shown in the planet's rest frame. For this case, the spots parameters are: \( f_{\text{sp}} \) = 1\%, $\theta_{lat}$ = $25^\circ$, $\theta_{lon}$ = $20^\circ$.}
\label{fig16}
\end{center}
\end{figure*}
\end{appendix} 
\end{document}